%Paper: hep-th/9307123
%From: tseytlin@surya11.cern.ch (Arkady Tseytlin)
%Date: Mon, 19 Jul 93 21:45:44 +0200
%Date (revised): Fri, 19 Nov 93 14:01:38 +0100

\input harvmac

%%%%%%%%%%%%%%%%%%%%%%%%%%%%%%%%%%%%%%%%%%%%%%%%%%%%%%%%%%%%%%
%%%%%%%%% PLEASE USE `B'-option of HARVMAC%%%%%%%%%%%%%%%%%%%%%%%%%%%
%%%%%%%%%%%%%%%%%%%%%%%%%%%%%%%%%%%%%%%%%%%%%%%%%%%%%%%%%%%%%%%%%%%%%%%%
% The paper has five figures appended in the second part as  a uuencoded
% tar compressed postscript file with  instructions
% how to unpack (call it figs.uu).  The resulting  5  fig*.ps files
% must be printed out separately.
%%%%%%%%%%%%%%%%%%%%%%%%%%%%%%%%%%%%%%%%%%%%%%%%%%%%%%%%%%%%%%%%%%%%%%%%
%  FILE --PLAIN TEX ----of paper  ``Charged string solutions ..."
%%%%     %%%%  by M. Cvetic  and  A.A. Tseytlin %%%%%%
%%%%%%%%%%%
%%%%%%%%%%     PLEASE ANSWER   "B" to query
%%%%%%%%%%%%%%%%%%%%%%%%%%%%%%%%%%%%%%%%%%%%%%%%%%
%     FIRST GOES SOME STANDARD INPUT FILE --harvmac
%%% and THEN PREPRINT ITSELF %%%%%%%%%%%%%%

%%%%%%%%%%%%%%%%%%%%%%%%%%%%%%%%%%%%%%%%%%%%%%%%%%%
% HARVMAC -- Please use `` B " option
%%%%%%%%%%%%%%%%%%%%%%%%%%%%%%%%%%%%%%%%%%%%%%%%%%%%
 %%%%%%%%%%%%%%%%%%%%%%%%%%%%%%%%%%%%%%%%%%%%%%%%%%%%

%%%%%%%%%%%%%%%%%%%%%%%%%%%%%%%%%%%%%%%%%%%%%%%
%%%%%%%%%%%%%%%%%%%%%%%%%%%%%%%%%%%%%%%%%%%%%%%%%% %%%
%%%%  FILE STARTS HERE
%%%%%
%%%%%%%%%%%%%%%%%%%%%%%%%%

\def \U { {\cal U}}

\def \V { {\cal V}}
\def \E { {\cal E}}
\def \P { \Phi }
\def \S { {\cal S}}

 \def \k1 {{1\over
k}}  \def \ov { \over }

\def \O {\Omega }

\def \ra {\rightarrow}

\def \a {\alpha}
\def \b {\beta}

\def \sh {{\rm sinh \  }}

\def \log {{\rm log \ }}
\def \ln {{\rm \ ln \  }}

\def \ch {{\rm cosh \  }}
\def \th {{\rm tanh \  }}

\def \1p {{1\over  \pi }}
\def \2p {{{1\over  2\pi }}}
\def \4p {{ {1\over 4 \pi }}}
\def \8p {{{1\over 8 \pi }}}
\def \p {\phi}

\def \m {\mu }
\def \n {\nu}

\def \k {\kappa }

\def \fourth {{\textstyle{1\over 4}}}

\def \e#1 {{{\rm e}^{#1}}}
\def \const {{\rm const }}

\def \eq#1 {\eqno {(#1)}}
%%%%%%%%%%%%%%%%%%%%%%%%%%%%%%%%%%%%%%%%%%%%

\def \O {{\cal O}}
\def \hr {{\hat r}}

\def \fourth {{1\over 4}}

\def \e#1 {{{\rm e}^{#1}}}
\def \const {{\rm const }}

\def \vp {\varphi}

\def\np {  Nucl. Phys. }
\def \pl { Phys. Lett. }
\def \mpl { Mod. Phys. Lett. }
\def \prl { Phys. Rev. Lett. }
\def \pr  { Phys. Rev. }

%%%%%%%%%%%%%%%%%%%%%%%%%%%%

\lref \ortin {T. Ortin, \pr {\bf D47} (1993) 3136.}
\lref \kaloper { N. Kaloper and K.A. Olive, University of Minnesota preprint
UMN-TH-1011, 1991. }
\lref \kallosh {R. Kallosh, A. Linde, T. Ortin, A. Peet and A. Van Proeyen, \pr
{\bf D46} (1992) 447.  }
\lref \derendi {J.P. Derendinger, L.E. Ib\~ anez and H.P. Nilles, \pl {\bf
B155}
(1985) 65. }
\lref \quiros {J. Garcia--Bellido  and M. Quir{\'o}s,  \np {\bf B385} (1992)
558. }
\lref \gibb {G.W. Gibbons and K. Maeda, \np {\bf B298} (1988) 741. }
\lref \ghs {D. Garfinkle, G.T. Horowitz and A. Strominger, \pr {\bf D43} (1991)
3140.  }
\lref \cas {  J.A. Casas, Z. Lalak, C. Mu{\~n}oz and G.G. Ross,
\np {\bf B347} (1990) 243;
 B. de Carlos, J.A. Casas and  C. Mu{\~n}oz,  \np {\bf B399} (1993) 623.}

\lref \horow {G.T. Horowitz, ``The dark side of string theory: black holes and
black strings", in
{\it Proceedings  of the 1992 Trieste Spring School on String Theory and
Quantum Gravity"}, preprint
UCSBTH-92-32. }
 \lref \maison {G. Lavrelashvili and D. Maison, \pl {\bf B295} (1992) 67. }

\lref \bizon {P. Bizon, \pr
{\bf D47} (1993) 1656. }

\lref \bhym {G. Lavrelashvili and D. Maison,  Munich preprint MPI-Ph/92-115;
P.
Bizon, University of Vienna  preprint ESI-1993-18 (1993); E.E. Donets and
 D.V. Galtsov, Moscow University preprint
DTP-MSU-93-01 (1993). }

 \lref \zumin {M.K. Gaillard and B. Zumino, \np {\bf B193} (1981) 221. }
\lref \dine {M. Dine, R. Rohm, N. Seiberg and E. Witten, \pl {\bf B156} (1985)
55. }
\lref \bartnic {R. Bartnik and J. McKinnon, \prl {\bf  61} (1988) 141; D.V.
Galtsov and M.S. Volkov, \pl
{\bf B273} (1991) 255. }
\lref \BrSt{ R. Brustein and P. Steinhardt,  \pl {\bf B302} (1993) 196.   }
\lref\derovr{G. Cardoso and   B. Ovrut, Nucl. Phys. {\bf B369} (1992) 351 ;
J.P.
 Derendinger, S. Ferrara, C. Kounnas and F. Zwirner,
Nucl. Phys. {\bf   B372} (1992) 145; J. Louis, in {\it Proceedings of the
International Symposium on
Particles, Strings and Cosmology}, Boston, March 1991,  P.
Nath and S. Reucroft eds. (World Scientific, 1992).}
\lref \cadoni {M. Cadoni and S. Mignemi, Cagliari Univ. preprint INFN-CA-10-93.
 }
\lref \tse { A.A. Tseytlin, ``String cosmology and dilaton",  in {\it
Proceedings of the 1992 Erice
workshop
 ``String Quantum Gravity  and Physics at the Planck scale"}, ed. N. Sanchez
(World
Scientific,1993).}
 \lref \stewart { N.R. Stewart, \mpl {\bf A7} (1992) 983.}
\lref \HH {J.H. Horne  and G.T. Horowitz,
\np {\bf B399} (1993) 169.}
\lref \GH { R. Gregory and J.A. Harvey, \pr {\bf D47} (1993) 2411. }
\lref \witten { E. Witten, \pl {\bf  B155} (1985) 151.
}
\lref\kap{V. Kaplunovsky, \np {\bf B307} (1988) 145.}
\lref \DKL {L. Dixon, V. Kaplunovsky and J. Louis, \np {\bf B355} (1991) 649.
}
\lref \shapere { A. Shapere, S. Trivedi and F. Wilczek, \mpl {\bf A6} (1991)
2677. }
\lref \font { A. Font, L. Ib\~anez, D. L\"ust and F. Quevedo, \pl {\bf B249}
(1990) 35;
 A. Sen,  Tata Institute preprint TIFR-TH-92-41 (1992); J. Schwarz,
 Caltech preprint CALT-68-1815 (1992);
J. Schwarz and A. Sen,  Santa Barbara Institute %for Theoretical Physics
 preprint NSF-ITP-93-46 (1993).}
\lref \ffont { A. Font, L. Ib\~anez, D. L\"ust and F. Quevedo, \pl {\bf B245}
(1990) 401;
\np {\bf B361} (1991) 194.}
\lref \fer {S. Ferrara, D. L\"ust, A. Shapere and S. Theisen, \pl {\bf  B225}
(1989) 363. }
\lref \anton {I. Antoniadis, J. Rizos and K. Tamvakis, Palaiseau
preprint CPTH-A239.0593 (1993). }
\lref \IbLu   { L. Ib\~ anez and D. L\" ust, \np {\bf B382  } (1992) 305.    }
\lref\CFILQ{M. Cveti\v c, A. Font, L. Ib\~anez, D. L\" ust and F. Quevedo
 \np {\bf B361} (1991) 194.}
\lref\shenker{S. Shenker, in {\it Proceedings of the Carg\` ese Workshop on
Random Surfaces, Quantum Gravity and Strings} (1990).}
\lref\brov{ R. Brustein and
B. Ovrut, Univ. of Pennsylvania preprint, UPR-523-T (1992).}
\lref\cqr{ M. Cveti\v c, F. Quevedo
and S.-J. Rey, \prl {\bf 63} (1991) 1836; E. Abraham  and P.
Townsend,  \np {\bf B351} (1991) 313;
M. Cveti\v c, S. Griffies and S.-J. Rey, \np {\bf   B381}  (1992) 301.}
\lref\khuri{  R. Khuri, \pr {\bf D46} (1992) 4526;
 J. Gauntlett, J. Harvey and J.T. Liu, University of Chicago preprint
EFI-92-67 (1992); M. Duff and R. Khuri, Texas A\& M preprint,  CTP-TAMU-17/93
(1993).}

\lref \ferr {S. Ferrara, N. Magnoli, T. Taylor and G. Veneziano, \pl {\bf B}245
(1990) 409.}
\lref \nill {P. Nilles and M. Olechowski, \pl {\bf B}248 (1990) 268. }
\lref\ctii{M. Cveti\v c and A.A. Tseytlin, UPR-589-T, in preparation.}

%%%%%%%%%%%%%%%%%%%%%%%%%%%%%%%%%%%%%%%%%
%%%%%%%%%%%%%%%%%%%%%%%%%%%%%%
\baselineskip6pt
\Title{\vbox
{\baselineskip4pt
\hbox{CERN-TH.6911/93}{\hbox{Imperial/TP/92-93/41}}{\hbox{UPR-573-T}}
{\hbox{hep-th/9307123}}
 }}
{\vbox{\centerline { Charged   string    solutions with}\vskip2pt
 \centerline{%non-trivial
dilaton and modulus fields  }
}}
\centerline {M.  Cveti\v c\footnote{$^{*}$}{\baselineskip8pt
e-mail: cvetic@cvetic.hep.upenn.edu }  }
\centerline {\it Physics Department}
\centerline {\it
 University of Pennsylvania, Philadelphia PA19104-6396, USA }
\medskip
\centerline { and}
\medskip
\centerline{   A.A. Tseytlin\footnote{$^{**}$}{\baselineskip8pt
On leave from Lebedev Physics
Institute, Moscow.  e-mail:
  tseytlin@surya3.cern.ch or tseytlin@ic.ac.uk} }
\centerline {\it Theory Division, CERN}
\centerline {\it
CH-1211 Geneva 23, Switzerland}
\vskip2pt
\centerline {\it  and }
\vskip2pt
\centerline{\it  Theoretical Physics Group }
\centerline {\it  Blackett Laboratory, Imperial College}
\centerline{\it  London SW7 2BZ, U.K. }
\medskip
\centerline {\bf Abstract}
\medskip
\baselineskip7pt
\noindent
We  find   charged, abelian, spherically  symmetric
solutions (in flat  space-time) corresponding to
 the  effective  action of $D=4$  heterotic string theory
 with the scale   dependent dilaton  $\p$ {\it and}
 modulus $\vp$ fields.  We  take  into account
 perturbative  (genus-one),  moduli-dependent `threshold'
corrections to  the coupling function  $f(\p,\vp)$ in
 the gauge field  kinetic term
$f(\p,\vp) F^2_{\m\n}$,  as well as non-perturbative  scalar potential
$V(\p, \vp)$, {\it e.g.},   induced by  gaugino
 condensation in the hidden gauge  sector.
  Stable, finite energy,  electric solutions (corresponding to on abelian
  subgroup of a non-abelian gauge group) have the small scale region as the
  weak coupling region ($\phi\rightarrow-\infty$) with the modulus $\vp$
   slowly varying towards  smaller values.
Stable, finite energy, abelian  magnetic solutions exist only for a specific
range of  threshold correction parameters. At small scales they
correspond to the  strong coupling region  ($\p\ra \infty$) and the
 compactification region ($\vp\ra 0$).
The non-perturbative potential $V$ plays a crucial role  at large  scales,
where
it fixes the asymptotic values of $\phi$ and $\vp$ to be at the minimum of $V$.
\smallskip
\noindent
{CERN-TH.6911/93}
\noindent

\Date {July 1993 }
%\draftmode
\noblackbox
\baselineskip 20pt plus 2pt minus 2pt

%%%%%%%%%%%%%%%%%%%%%%%%%%%%%%%%%%%%%%%%%%%%%%%%%%%%%%
%%%%%%%%%%%%%%%%%%%%%%%%%%%%%%%%%%%%%%%%%
%%%%%%%%%%%%%%%%%%%%%%%%%%%%%%
%%%%%%%%%%%%%%%%%%%%%%%%%%%%%%%%%%%%%%%%%%%%%%%%%%%%
\newsec {Introduction }
One of the basic  features  of string theory is the existence of
scalar fields, such as dilaton and moduli. The latter
ones are string modes  in a vacuum  associated with
 compactification of extra dimensions. In the simplest cases moduli correspond
to  the `radii' of compact
dimensions. These fields are natural partners of the metric
and thus should play an important role in string gravitational physics.

Moreover,  it is  a generic property
of  Kaluza-Klein-type  theories, and thus also of string theory,
that scalar  fields  couple   to the   Maxwell
and  Yang-Mills  kinetic terms of the gauge fields.  In particular,
in string theory,  the dilaton field determines the strength  of the gauge
couplings at the tree level of the effective action, while  string one-loop
(genus-one) contributions   give  \DKL\ moduli dependent corrections to such
couplings.
Thus, in general,   a scalar function  $f$ that couples to  the gauge field
kinetic energy is  a function of both the dilaton as well as the moduli.

The dilaton and  the moduli have no potential in the effective action  to all
orders in
string loops.  To avoid a  contradiction with observations
they should  acquire masses.  Currently proposed scenarios
rely on  non-perturbatively  induced potential $V$ due to gaugino condensation
in the hidden gauge group sector.\foot{%
Other non-perturbative string effects may lead to contributions to the
potential $V$ which have different functional dependencies on the dilaton
$S$, {\it e.g.,} $V\sim e^{-\sqrt S}$ \shenker. See also \brov\ and references
therein.}
 Such a potential would  generate masses
for the dilaton and the moduli
 and  at the same time provide a  mechanism   of  supersymmetry breaking.
 While these  scalar fields    eventually get masses,
 it is very important to appreciate the fact
that they may change with distance (or time, in the cosmological context) at
small scales (or times).
In other words,  such  fields may participate in the dynamics at
 small scales (or times),
 {\it i.e.}, in the region  where  non-perturbative effects presumably can be
neglected.

%%%%%%%%%%%%%%%%%%%%%%%%%%%%%%
String solutions are usually   discussed in
perturbation theory in $\a'$ (string tension); occasionally it is possible to
go beyond the $\a'$
expansion by identifying an exact conformal field theory which corresponds to a
given
tree-level solution.
For example,  new   charged black hole  string  solutions \gibb\ghs\ have been
recently obtained   by
taking into  account  the  tree level coupling of the dilaton to the
 gauge fields (for a review see \horow).
However, there are very few discussions in the literature where
perturbative ({\it e.g.},  genus-one,  moduli dependent threshold corrections
to the
gauge couplings),  or non-perturbative  string effects
 ({\it e.g.}, non-perturbatively induced potential
for the dilaton and moduli)  are taken into account.\foot{
There were discussions of cosmological solutions in the presence of a
non-perturbative
 dilaton potential
(see e.g. \kaloper\quiros\tse\stewart\BrSt\ and refs. therein).
Aspects of charged dilatonic black hole solutions with non-perturbatively
induced dilaton
mass included
 were addressed  in \GH\HH.
 However, in the latter case,
 the potentials for the dilaton were  not always taken
 to be `realistic' or well motivated from the point of view of
 non-perturbative dynamics like gaugino condensation in the hidden gauge
 sector of
the gauge group.}
Such corrections may substantially modify the tree level solutions and
it  is thus important to include
them in order to understand  predictions of string theory.

The change of the gauge coupling function from the
naive exponential dilaton factor  in the three level action to a more
realistic one, as found in string perturbation theory, may have a fundamental
influence on the
previously discussed solutions.  Addressing the case of a general gauge
coupling
was one  of the  motivations for the present  paper.
One should thus
consider the following improvements of  the previous  approaches:
{\it (i)} A more general form of the dilaton potential should be taken into
account.
Up to now,  solutions with the  dilatonic mass term \HH\  and/or
 a  choice  of  {\it convex} potential \GH\
were studied.  One should  analyse    subtleties associated with the fact
that starting from the weak coupling region
the `gaugino  condensation'  potential always has a
concave region, {\it i.e.}, the potential grows as the coupling
increases.
{\it (ii)} The dynamics of the modulus field was ignored. The modulus,
however, also couples to
the gauge field strength  due to the one-loop threshold corrections. It also
appears,
along with the dilaton, in the non-perturbatively induced  potential.
Therefore, one  may not  {\it  a priori}  set  any of the two fields  equal
 to a constant.
{\it (iii)} A possibility of additional non-perturbative dilatonic terms in $f$
was not considered.

The present work is a step  towards   clarifying  some of  these issues.
In this paper, however, we
shall  ignore the  gravitational dynamics, thus  postponing a study of
black hole--type  solutions  for future work. We shall
consider      abelian electric  and magnetic solutions
 in flat space concentrating on the role of the non-trivial functions
$f$, a  coupling function of scalars  to the gauge field kinetic  term, and
 $V$, a non-perturbative potential for the dilaton and the moduli.

The solutions we shall find  should have generalisations to the curved space.
We shall assume that they approximate the exact solutions  of the whole set
of equations (including the gravitational one) in the region where the
curvature is small.
As in the case of solitonic solutions in  field theory in flat space  one can
ignore
graviational effects  if the  scale of the solutions is  large  compared to the
 gravitational scale ($\sim E/M_{Pl}^2$)
(i.e.  if the  energy of the solutions is small enough).
It is true that in the absense of  a non-perturbative potential the dilaton and
the metric
are on an equal footing. Once the potential is generated, it introduces a new
scale
(different from the Planck one). This   makes  it possible  in principle
to  `disentangle' the metric from the dilaton  and to consider  dilatonic
solutions
in flat space  with a characteristic scale  being larger than   the
graviational one.

In our study we shall
include the dilaton $\p$  and, for the sake of simplicity, only one    modulus
field $\vp$,  which is
associated with  an overall compactification scale.
We shall look  for stable spherically symmetric
finite-energy  solutions  with  a  regular  gauge  field
strength in flat $D=4$ space-time.
We shall  consider a
general class of functions $f$ and $V$.  We will treat examples
with $f$ modified by the string loop corrections, as they appear \DKL\ in a
class of orbifold-type  compactifications, and with the
non-perturbative  potential
$V$  due to the gaugino condensation in the hidden sector of the theory,
as special cases.
 We shall not include  higher derivative terms, assuming that  the
fields change slowly in space.

We shall find that the abelian electric solutions\foot{
Note that  regular, finite energy  spherically symmetric solutions
 exist in the tree-level  ($f=\e{-2\p} , \
V=0$)  dilaton-Yang-Mills  system,  but
  are, in general,   unstable \maison\bizon\
(they are similar to the regular  but unstable solutions
of the Einstein-Yang-Mills system
found in \bartnic). The simplest abelian solutions of this class  are,
in fact, stable
and are the obvious  limiting cases  of the gravitational
solutions of \gibb\ghs.}  are regular, have finite energy,
and  are stable when the abelian subgroup is
 embedded in a non-abelian gauge group.
They  have the  effective string coupling $\e{\p} $  increasing from  zero
at the origin ($r=0$) to a finite value $\e{\p_0} $ at $r=\infty$. The
asymptotic value  $\p_0$ of the dilaton
corresponds to the minimum of the potential $V$.
Thus the small  distance region is a weak coupling region  and
can be studied ignoring non-perturbative corrections. The   large distance
region
corresponds to  the `observed' world where the dilaton is trapped in the
minimum
of $V$.
 The modulus field $\vp$ is slowly  varying with $r$; at large scales it is
fixed
at  the minimum of the potential $V$, while at small scales its value
decreases slightly. Generic  existence of such   particle-like, finite energy,
  charged    configurations   may have
potentially interesting applications.

Stable, regular, finite energy, abelian magnetic solutions   exist for a
certain
range of threshold correction parameters.
 Here at small distances   ($r\ra 0$) the  dilaton  approaches
the  strong coupling region, $\e{\p} \ra  \infty$, while
 the modulus  goes to zero, $\vp\ra 0$, {\it i.e.},
  the small scale region is the compactification region.
The role of the non-perturbative potential is again to fix the asymptotic
values of $\p_0$ and $\vp_0$ to be at its  minimuml.

The plan of this paper is the following.
In Section 2 we  discuss  the string effective action with the dilaton and the
modulus field and its modification due to the
perturbative  as well as non-perturbative corrections. In Section 3 we
study  general properties of the solutions
without specifying particular  functions $f$ and $V$.
In Section 4 we
turn to specific examples (with only one of the two fields changing with $r$)
with
particular choices for the function $f$, but ignoring the  potential ($V=0$).
 In Section 5 we   consider  the
case when  both the dilaton and the modulus field  are included in $f$, while
the potential is still zero.
The modifications of the solutions due to the
non-perturbative potential are considered in Section 6.
Section 7  contains  a summary and  concluding remarks.

\newsec {Structure of low-energy string effective action}

\subsec{Perturbative terms in the effective action}

The leading terms (in the derivative expansion) in the
low-energy, $D=4$ effective action of the heterotic string theory
have the form\foot{We use the  space-time signature $(- + + +)$ and the
gravitational constant $G=1$.}
$$S=
{1 \ov 16 \pi }
\int d^4 x {\sqrt {-g }}\ \left[ R - 2 \del_\mu \p\del^\mu \p - 2 \del_\mu
\vp\del^\mu\vp - f(\p, \vp) F_{\m \n} F^{\m\n}
 -  V (\p , \vp) + ...  \right] \ \ . \eq{2.1} $$
Here  $g=|det g_{\mu\nu}|$, $R$ is the scalar curvature and  $F_{\mu\nu}$ is
the
abelian gauge field strength. For simplicity we consider,   along with
the dilaton field $\p$,  only
 one modulus field $\vp$, associated
 with an overall scale of compactification and  ignore in the most part of
 the paper the axionic
partners $(\a, \b)$ of $\p$ and $\vp$ as well as other matter fields.
The standard scalar fields $(S,T)$ of  chiral multiplets
of $N=1$ supergravity,  corresponding to  the
dilaton and the modulus are \witten\
$$ S= \e{-2\p}  + i\a \ , \ \ \ \ T= \e{2\vp /\sqrt 3}  + i\b \
.\eq{2.2} $$
%In what follows we shall often set $\a=\b=0$.
At the tree level
$$ f_{tree}= \e{-2\p} \ \ , \ \ \  \ \  V_{tree}=\  c \ \e{2\p} = 0 \ ,
\eq{2.3} $$
where $c$ is the  central charge deficit.

 In the case  of the supersymmetric $D=4$ heterotic string $c=0$
  and  thus $V$ remains zero to all orders in the string perturbation theory.
On the other hand,
the gauge coupling function $f$ receives a non-trivial, $\vp$-dependent, string
one-loop  (genus-one) correction \DKL. Thus:
$$ f_{perturb} = \e{-2\p} + f_2(\vp) \  , \ \ \ \  V_{perturb} =0 \ . \eq{2.4}
$$
The modulus dependent function
$f_2$  depends on a type of superstring vacuum one is considering. In
particular, for toroidal compactifications and  a class of orbifolds, it
is invariant under the duality  symmetry ($\vp \to -\vp$) and
 can be schematically written in the form:
 $$ f_2(\vp)  = b_0 \ln\left[(T + T^*) |\eta (T)|^4 \right] +b_1,\ \ \
T=\e{2\vp/\sqrt 3} \ \  .
\eq{2.5} $$
Here $\eta (T)$ is the Dedekind function (modular function of
weight $-1/2$). It turns out that  $\ln  \left[(T + T^*) |\eta (T)|^4\right]$
is always negative,  has a maximum at $\vp=0\,(T=1)$ and approaches
$ - {\pi\ov 3}T =-{\pi\ov 3}\e{2\vp/\sqrt 3}
$
 as  $\vp \ra \infty$. The important property
of  $f_2$ is its duality symmetry $f_2(\vp) = f_2(-\vp)$. The constant $b_0$
is related to the one-loop $\b$-function coefficients associated with the $N=2$
subsector of the massless
spectrum in a symmetric orbifold compactification.\foot{The coefficient
$b_0$ contains in general  a contribution  due to the mixed
Yang-Mills--sigma model anomaly \derovr\ .
 In  special cases, {\it i.e.}, of $Z_3$,  $Z_7$
orbifolds, the total value of $b_0$  turns out to  be zero. In most
cases, however, {\it e.g.}, $Z_4$, $Z_6$ etc. orbifolds,
 $f_2$ depends  on the contribution of a  modulus associated with  one
two-torus and not on
a modulus associated with an overall scale of six-torus.
In the following, we consider  a symmetric contribution to $f_2$ which
depends  on an overall modulus.  A study with a modulus associated  with
one two-torus only  can be done   analogously and should
 yield the same qualitative
features.}
Generically
$b_0=\O (1/100)$  and  is negative (positive) in the case of  the abelian
(non-abelian)
 gauge  group factors, although examples with
reversed signs were also found \IbLu.
There is also  a constant, moduli-independent and non-universal
contribution \kap\ to the gauge coupling threshold correction term,
 denoted  as $b_1$ in eq. (2.5). According to \kap\
  $b_1={\cal O}(1/100)$ and it is  positive
(negative) for the abelian (non-abelian) case.
Thus, in general $f_2(\vp)$ is positive (negative) in the
abelian (non-abelian) case. In Figure 1 the function
$f_2(\vp)$ with $b_0=-1$ and
$b_1=0$ is plotted.

%%%%%%%%%%%%

Both $f$ and $V$ may, in principle,  receive  as well corrections which are
non-perturbative
 in string coupling.
Since the string  coupling is related to the dilaton,
both $f$ and $V$  may contain non-perturbative contributions
which are  non-trivial functions of $\p$.
In fact, {\it  a priori}  separate  non-perturbative factors may appear
in  the  kinetic term of the
dilaton, in the gauge  field term and in the potential so that  after a field
 redefinition
not only $V $,  but also  $f$ may  contain non-trivial dilaton dependence.
Such terms in $f$ do not actually appear in
the proposed gaugino condensation scenario for
 supersymmetry breaking. One should, however,  bear in mind,
that the origin of supersymmetry breaking in string theory is
not well understood. In view of that  one should allow for
a possibility  that $f$ may contain additional, non-perturbative
terms which depend on $\p$, {\it e.g.},   $\exp{[- k
\exp (-2\p)]} $ (implying   $ 1/g^2 \ra 1/g^2 \  + \ a \  \e{ - k/g^2}  $).

%%%%%%%%%%%%%%%%%%%%%%%%%%%%%%%%%%%%%%%%%%%%%%%%%%

\subsec { Non-perturbative scalar potential }
%%%%%%%%%%%%%%%%%%%%%%%%%%%%%%%%%%%%%%%%%%%%%%%%%%%
We now summarize the properties of the  non-perturbative
  potential for the dilaton and the moduli fields.
A detailed  structure of a non-perturbative potential $V(\p, \vp)$
depends  on a  particular mechanism of
supersymmetry breaking.  We shall  describe the form of
$V$ due to the gaugino condensation in the hidden  sector
of the  gauge group \derendi\dine\ffont\ferr\nill\DKL\cas.

The $N=1$ supergravity potential can be written in terms of a K\" ahler
function  $${\cal G}=K+\ln |W|^2 \ , $$,where $K$ is the K\" ahler potential,
and $W$ is the
superpotential. In the case of gaugino condensation $\cal G$ is a separable
function, {\it i.e.},
$${\cal G}={\cal G}_1(S,S^*) +{\cal G}_2(T,T^*)\ \ \ , \eq{2.6} $$
where  ${\cal G}_1(S, S^*)$
depends on  the dilaton $S$,  and ${\cal G}_2(T, T^*)$ depends on the modulus
$T$ only.
The potential is then of the form:
$$V(S,T)=\e{{\cal G}}  \left({\cal G}_{SS^*}^{-1}|{\cal {G}}_S|^2+{\cal
G}_{TT^*}^{-1}|{\cal G}_T|^2-3\right)\
\ \ ,
\eq{2.7} $$
where ${\cal G}_S={\del {\cal G}/  \del S}$, $\ {\cal G}={\del^2
{\cal G}/  \del S\del S^*}$, etc.
In the case of  symmetric  orbifolds  with  the
compactification moduli
of all three two-tori equal to $T$ the K\" ahler functions are of the
following form \ffont\ferr\nill:
$${\cal  G}_1(S,S^*)=-\ln (S+S^*)+ \ln |H(S)|^2, \ \ \
{\cal G}_2(T,T^*)=-3\ln (T+T^*) -6\ln
|\eta(T)|^2  \ \ .\eq{2.8} $$
Here
 $$H(S) =
\sum^J_{i=1} d_i \e{-a_i S}  \ , \  \ \ a_i= {3\ov 2b_{0i}}\ ,
\eq{2.9} $$
where $J$ is a number of gaugino condensates and $b_{0i}$ are
the (one-loop $N=1$)
$\b$-functions
of the gauge group factors  of the hidden gauge group sector.
 Inserting (2.8) in the potential (2.7) one finds \ffont\ferr\nill\foot{For
general superpotentials generating duality invariant potentials
see \CFILQ .}
$$   V_{non-perturb.}\equiv V(S,T)
 ={|H|^2 \ov |\eta (T)|^{12} S_RT^3_R } \left[ |S_R {{\del \ln H} \ov \del S }-
1  |^2
 + {3\ov 4\pi^2}     T^2_R    |{\hat G}_2(T)|^2 -3   \right] \ , \eq{2.10} $$
where $S_R = 2 {\rm Re} S ,  \ T_R= 2 {\rm Re} T $ ,  and
 $$\ {\hat G}_2(T) =  G_2 (T) -2\pi T^{-1}_R=  - 4 \pi {1\ov \eta (T)} {\del
\eta (T)  \ov \del
T } - 2\pi T^{-1}_R \  $$
is the  Eisenstein function of weight $2$. It has zeros at $T=1$ and
$T=\e{i\pi /6} $, the respective  $Z_2$  and $Z_3$ symmetric points of the
fundamental domain
of the $SL(2,{\bf Z})$ modular group.

This potential  vanishes in the weak
coupling limit $S = \e{-2\p} \ra \infty $ and has an extremum in
$S$ if  $ S_R {\del W\ov \del S }- W
=0$.  A minimum exists if $J >1$, {\it i.e.},
in cases with more than one gaugino
condensate.
Since the condition for a minimum in $S$ is
 ${\cal G}_S= H^{-1} {\del H/ \del S }- S_R^{-1}  =0$,
 the dilaton sector does not break supersymmetry.
 As for the extrema in $T$,   $\del V/ \del T = 0$, they are achieved
  at the self-dual points
  $T=1$ and
$T={\rm e}^{i\pi /6}$,
which are saddle points of $V$, and at $T\sim 1.2$, which is
the  minimum of $V$.  Interestingly,
  both   $f$ and $V$    have an extremum in  $T$  at the points   $T=1$
   and $T=\e{i\pi /6} $ (zeros of ${\hat G}_2$),    since  $ {\del f /\del T }
 = - {b_0 \ov \pi }{ \hat G}_2$ and
 ${\del V/\del T }\propto { \hat G}_2  \ $ (see \fer\ffont\ferr\nill\CFILQ).
At these points $V$ preserves supersymmetry in the
$T$ sector (note that ${\cal G}_T\propto {\hat G}_2$).
At  a fixed,  extremal value of $T$  and  a fixed ${\rm Im} S$,\foot{In order
to
have a  minimum of  $V$ for a finite value of  ${\rm Re} S $,
 in such a minimum  one should have, in general,
  ${\rm Im}  S\ne 0$.}
the potential $V$ can be represented in the following form
$$ V= S_R^{-1} \sum^J_{i=1} {\rm e}^{-a_i S_R} (c_i  + d_i S_R + e_i S_R^2) \ ,
 \eq{2.11} $$ where $a_i$, $c_i$, $d_i$, and $e_i$ are constants
and $S_R = 2 {\rm e}^{-2\p}$.
  For example, in  the case of two gaugino condensates,
     $J=2$,  this
potential starts from zero in the  weak coupling region $\p \ra -\infty$,
grows  and reaches a local
maximum, then  decreases to a local minimum (with negative value of $V$),
then  has the second  local
maximum and finally  goes to $-\infty$ at   $\p \ra +\infty$. Since the
potential  has a local minimum in $\phi$, it may
fix the value of the dilaton.\foot{This would
 give  a mass  to the fluctuating part
of the dilaton.
A generic  property of this potential is that starting from a weak coupling
region it first
increases and has a local maximum and only then decreases to a minimum, namely,
the potential is {\it not} convex everywhere.
 Another problem is that the value of the potential at the
minimum, {\it i.e.} the effective cosmological constant, is negative in
general.
A local minimum with a zero cosmological constant can be achieved in the case
 with
  more than two gaugino condensates.
Note, however, that in this case there
are usually also other minima with negative cosmological constants (see {\it
e.g.}, \BrSt).}

%%%%%%%%%%%%%%%%%%%%%%
\newsec{ General  properties of  charged solutions in flat space-time  }
\subsec{Equations of motion} In the following we shall study
 charged solutions with  non-trivial
dilaton $\p$  and  the modulus field $\vp$
in flat four dimensional  ($D=4$)  space-time, {\it i.e.},
$ds^2=-dt^2 + dr^2 + r^2
d\Omega_3^2$. We shall look
 for   spherically symmetric,   static solutions  with
non-trivial abelian gauge field strength. The same approach
applies also to solutions associated with an abelian subgroup of
a  non-abelian gauge group. We shall refer to  such solutions as  abelian
solutions
embedded in  a non-abelian theory.\foot{We are making this distinction
because of the opposite signs of the coefficient $b_0$ in $f$ (2.5) in the
abelian and non-abelian cases. }

The  relevant part of the action (2.1)  has the form\foot {For convenience we
rescale the potential $V$ by the factor 4 as compared to $V$ in  (2.1).}
  $$S= {1\ov  4   \pi } \int d^4 x \
\{  - \ha   (\del \P_i)^2 -  \fourth f(\P_i) F_{\m \n} F^{\m\n}
 -  V (\P_i) \} \ \ ,  \eq{3.1} $$
so that the corresponding  field equations  are
 $$ D_\m (f F^{\m\n}) =0 \ ,\ \ \  D_\m  F^{*\m\n}
=0 \ , \eq{3.2} $$ $$ D^2 \P_i -  \fourth  \del_i f  F_{\m\n} F^{\m\n} - \del_i
V = 0 \ , \ \ \ \
i=1,2  \ , \eq{3.3} $$
 where  $ \del_i= {\del/\del \P_i} $ and $ \P_i = (\p , \vp )
$. It is easy to see that this system  transforms into the same  one  under
 the  following  `duality
transformation' (see, {\it  e.g.},  \zumin )
 $$ f\ra  f^{-1} \  , \ \ \
  F_{\m\n} \ra f  F^*_{\m\n} \ , \ \  \P_i\ra\P_i \ .$$
In particular,  this implies that  electric solutions  for the action (3.1)
with
the  gauge coupling function $f(\P_i)  $ are   related to the magnetic
solutions of the  theory with
the coupling  $f^{-1}(\P_i) $.

 Eqs. (3.2--3.3)  may have also  standard symmetries (for fixed
functions $f,V$).  For example, if  the modulus can be ignored,  {\it i.e.}, in
 the
pure dilaton case  ($\P_i=\p$,   $f= e^{-2 \p}$ and $V=0$), eqs. (3.2--3.3)
are
invariant under the usual duality transformation  of $F$ combined
with $\p\ra -\p$. Such a  symmetry may survive also if $V(\p)=V(-\p)$.

 In the case   when the theory is invariant under  the
 modulus   duality symmetry $\vp\ra -\vp$
(as  happens  for  toroidal as well as a class of orbifold
compactifications)  both $f$ and $V$  are invariant under
this  duality,  which  is thus  a symmetry of (3.2--3.3).

 Let us consider  first the electric solution:
$$  F_{01}= E(r) \ ,  \ \  \ F_{0k}=F_{1k}=F_{kl}=0\ (k,l=2,3) \ , \ \ \ \
 \P_i = \P_i (r) \ .
$$
Then  eq. (3.2) reduces to
$$  {d\ov dr} ( r^2 f E) = 0 \ \ ,
 \ \eq{3.4} $$
which has an obvious solution
$$ E=  {q\over { r^2 f(\P_i (r))}} \ .\ \eq{3.5} $$
Using (3.5), one simplifies  the form of eqs. (3.3) to:
$$ \P_i'' +  \del_i U -  {1\ov x^4 } \del_i V = 0   \ ,  \ \ i=1,2 \ ,
 \eq{3.6} $$
where
$$  U\equiv -{q^2\ov {2f}}\ .$$
We have introduced a new coordinate
$ x = {1/ r} \ $ with the range  $  0\le x \le \infty $,  and
$\P_i'\equiv d \P_i / dx  $.
% In addition  the ``potential'' $ U (\P_i) \equiv -{q^2\ov 2 f }  \   $.

Equation  (3.6) can be interpreted as corresponding to a  mechanical system
with the action
$$ \S = \int^\infty_0  dx  ( \ha  \P_i'^2    + {q^2\ov 2  f }
 +  {1\ov x^4 }  V
)
 \ ,  \eq{3.7} $$
which,  at  the same time, gives the energy  of
 the  field configuration  as  derived from the action (3.1):
$$ \E = \S =
 \int^\infty_0  dr \ r^2  \ [ \
 \ha  ({d\ov dr} \P_i)^2
 +   {q^2\ov  2r^4  f }
  +   V \ ] =
\int^\infty_0  dx  ( \ha  \P_i'^2
 - U
 + {1\ov x^4 }  V ) \ .  \eq{3.8} $$
In eqs. (3.7--3.8) the summation over the index $i$ is implied.
Note, that the case with $V=0$ corresponds to the mechanical system with the
conservative potential $U$, while the case with $V\neq 0$
corresponds to the mechanical system with   non-conservative
(time-dependent) potential
$$\U (\P_i, x)\equiv  U -  {1\ov x^4 } V \ .$$

\subsec{Stability Constraints} Let us now discuss the stability constraints for
the  solutions
 of eqs. (3.6)  under
linearized perturbations. Expanding the action (3.1) around the solutions
of eqs. (3.6)  we find  for the terms quadratic in perturbations $\eta_i$
 and $B_\m$ of the
scalar fields   and the vector potential, respectively:
 $$S= {1\ov  4   \pi } \int d^4 x \
\{  - \ha   (\del \eta_i)^2 -  {1\ov 8} \del_i \del_j f \eta^i \eta^j F^2_{\m
\n} (A)  +   \ha f(\P_i) F^2_{01} (B) + \del_i f  \eta^i F_{01}(A) F_{01}(B)
$$ $$ -     \fourth  f F^2_{rs} (B)
 -  \ha \del_i \del_j V  \eta^i \eta^j  + ... \} \ \ . \eq{3.9} $$
Here $r,s=1,2,3$ and the summation over $(i,j)$ is implied.
Since $f $ is assumed to be positive the terms
which are quadratic in $B_\m$
give positive contributions to the energy and hence do not produce
instability.
%`Diagonalizing' (3.8)
% by integrating  over $B_\m$, or, equivalently,
Eliminating
$F_{\m \n}$   from (3.9) and using (3.5-3.6)
we get for the perturbations of the scalar fields
 ${ \eta_i} = r {\tilde\eta}_i (t,r)$
 $$S= \int dt \int^\infty_0  dr    \
\{  \ha   (\del_t {\tilde\eta}_i)^2 -
\ha (\del_r {\tilde\eta}_i)^2
           -  {1\ov 2 r^4}  \V_{ij}   {\tilde\eta}^i {\tilde\eta}^j
            + ... \} \ \ , \eq{3.10} $$
where
  $$  \V_{ij}  = - \del_i \del_j U   +
r^4 \del_i \del_j  V  = -  \del_i \del_j \U  \ . \eq{3.11} $$
A sufficient  condition of
 `linearized' stability  of  the abelian  electric solutions
is  therefore
 that  the matrix (3.11) should  have  non-negative  eigenvalues for all
$r$:
$$\V_{ij}\ge 0 \ . \eq{3.12} $$
 This condition has  an obvious interpretation that
perturbations should not decrease the energy (3.8) of the system.
 In the case when $V=0$ (3.12) is satisfied if $U$ is convex  within  the range
of variation of
$\P_i$.

Note, however,  that while the condition $\V_{ij}\ge 0$ is sufficient
 it  may  not be necessary.
Depending  on the nature of functions $f$ and $V$ the corresponding
Laplace operator may  have  only  non-negative  eigenvalues even
if (3.12) is not  satisfied. In other words, the
corresponding Laplace operator with the potential $\V$
need not have bound states with negative energy.

The set of equations, the energy  and the condition of stability for the
magnetic  solution
 $$F_{23}= h  \sin \theta   \ ,  \ \  \  \  h= {\rm const.} \  , \ \ \
 F_{01}=F_{0k}=F_{1k}=0\  , \ \
 k=2,3 \ ,  \ \  \ \ \P_i = \P_i (r) \ , $$
are found  by replacing $f$ by $f^{-1}$  and $q$ by $h$ in the
above equations or  by using
 $$ U\equiv -{1\ov 2} {h^2f}\ \ .$$
 %When $V=0$ the equations can be
%interpreted as  equations of a  mechanical system with
%the potential while with $V\neq 0$  it
% eqs.(3.12)
%corresponds to a non-conservative mechanical system with  $\U=U+{V\ov x^4}$.
 %{\it i.e.}, by using the duality symmetry of the eqs.(3.1)
% in (2.6),(2.7),(2.8),
%(2.11) and (2.15), i.e.
%with the energy
%$$ \E= \int^\infty_0  dx  ( \ha  \P_i'^2  + \ha {h^2  f }
% +  {1\ov x^4 }  V ) \ ,  \eq{3.12} $$
%Analogously, the sufficient  stability constraints can be rewritten as:
%$$  \del_i \del_j  f   +   {2r^4 \ov  h^2} \del_i  \del_j  V  \geq 0   \ .
%\eq{3.13} $$
%Thus, for $V=0$, the function $f$ has to be convex in the whole region of the
%of the $\P_i$.
Having spelled out  the  basic formalism, we would now like to proceed  with
 properties of examples.
\newsec { Case of zero scalar potential: examples of solutions with one
non-trivial
scalar field}
\subsec {Basic  relations }
It is clear that the
properties of the solutions  are different depending     on whether $V$ is
zero or not: in  the former case the system is conservative, in the
latter it is not.  In this Section and Section 5  we shall consider
  the case  when the effect
of the potential can be ignored, {\it i.e.},   when non-perturbative
corrections are small. Thus, we shall put $V=0$.
 In addition, in this section we address the properties of the solutions
with only one non-trivial field $\P_i\equiv \P$ for a  class of
functions $f(\P)$.

 This corresponds to the case when only one
scalar  field  $\P$   is
changing  with $r$.
Such a reduction to only one  field
 could be  possible if  $f$  had extrema with respect to
the other field and thus the other field
%in other directions so that they
 could  be `frozen', {\it i.e.}, put to a fixed $r$-independent value.
 In principle, $\P$  may be either the dilaton or  the  modulus,
but  as we shall see  later it does not seem to be possible  to
`freeze out' the $r$ dependence of the dilaton in general.

%Again, for concreteness we discuss the electric solutions.
If $V=0$   eq. (3.6)   has one integral of motion , {\it i.e.}, the
`energy' of the corresponding mechanical system:
$$    \ha  \P'^2  +  U(\P) =
 C=\const  . \eq{4.1} $$
The energy  (3.8) then  takes the form $$ \E =
 \int^\infty_0  dx  (  \P'^2  -  C) \ ,
  \eq{4.2} $$
where $U=-{1\ov 2}q^2f^{-1}$ and $U=-{1\ov 2} h^2f$ for the electric
 and the magnetic solutions, respectively. Here $x=1/r$ and $\P'= d \P/ dx$,
 as before.  In most cases,  for  finite energy solutions  one has to set
$C=0$.
We shall, however,   consider  also examples of  finite energy solutions
with $C\ne 0$.

In what follows we shall assume  that  $f$ is always positive (since it is the
coupling function in the gauge field kinetic term) and thus,
$U$ is  always negative.
 Then the absolute minimum of the
energy  (3.8)  is  achieved when  $\P$  takes the values corresponding
to  the  maximum of $U$ if it   exists.  In looking for other, $x$ dependent,
minima
of $\E$ it is useful to represent
 (3.8)  in the form
$$ \E =
\int^\infty_0  dx  \  \ha  \left( \P' \pm
 \sqrt {2C-2U(\P)}\right) ^2 \mp
\int^{\P_\infty}_{\P_0} d\P\sqrt{2C-2U(\P)}-
\int_0^\infty dx C  \ ,   \eq{4.3}
$$
where ${\P_\infty}\equiv\P(x=\infty)$ and $
 {\P_0}\equiv\P (x=0)  $. Note that $x\equiv {1/ r}$.
 In the following we shall consider solutions  with a regular field value,
{\it i.e.}, $\P_0\ne\infty$
 at $x=0$. However, in the core of the solution, {\it i.e.}, at $x\ra\infty$
the field may or may not blow up.

 The first  term  in (4.3) is always  positive  and hence the energy is
minimized  if $$  \P' =\mp\sqrt{2C-2U(\P)}\ \eq{4.4} $$
is satisfied.  The choice of sign is such that the second  and the third term
in eq. (4.3)
yield a  positive number. In addition,  the  constant $C$ is chosen such
that the  second and the third term in eq. (4.3) give a finite value of the
energy.
Eqs. (4.3--4.4) are  analogs of the
Bogomol'nyi-type  equations.\foot{Solutions satisfying eq. (4.4) extremize the
energy (4.3) for any value of $C$. However,  only
for a special value of $C$ (in most cases $C=0$) the solution has  a finite
energy. One has traded one of the two boundary conditions for $\P$ for the
constant $C$.
Note also that
eqs. (4.3--4.4) bear close similarities to analogous supersymmetric
systems, {\it e.g.},  supersymmetric walls \cqr.
Here the role of the superpotential is played by $W=\int_{\P_0}^{\P(x)}
d\P\sqrt{2C-2U(\P)}$. See also  Section 5.2.}

Equation (4.4) implies  the explicit form of solution
$$\int^{\P (x) }_{\P_0}
d\P {1\ov {\sqrt{2C-2U(\P)}}}   = \mp  x \ . \eq{4.5}   $$
 The upper (lower) sign solutions  correspond to $\P$
decreasing  (increasing) with increasing
$x={1/ r}$.
Then  the  energy $\E$ of the solution
is given by  $$  \E =
\mp  \int^{\P_\infty}_{\P_0}d\P \sqrt{2C-2U(\P)}
-\int_0^\infty dx C\ . \
\eq{4.6} $$
For the  electric solution the charge  $Q$
and  the  scalar charge $D$ are $$Q= [r^2 E(r)]_{r\ra \infty}  = {q\ov f(\P_0)}
,
\ \ \
  D = -[r^2 {d \P \ov dr }]_{r\ra \infty}=
%\P' (x=0) =
\mp
\sqrt{2C+{q^2\ov f(\P_0)}} \ . \eq{4.7} $$
The magnetic solution  is found from the electric solution
by the replacements $f\ra f^{-1}$ and $q\ra h$, or in other words, by taking
$U=-{1\ov 2} h^2f$.
 The expression for the energy (4.6)  suggests that
 generically for a given $f$
finite energy electric and magnetic solutions
correspond to  different values of $C$ as well as opposite signs in eqs.
(4.5--4.7).

In order to  have  a finite energy,
 regular solution the potential
 $U(\P)=-{1\ov2}{q^2 f^{-1}(\P)}$ (for the electric solution) and
 $U(\P)=-{1\ov2}{h^2f(\P)}$
 (for the magnetic solution) should satisfy
certain conditions.
The nature of the solution is different in the cases with $C=0$ or $C\ne 0$.
%In particular, one is searching  for  regular, finite energy configurations.
 One can show that the regular, finite energy  solutions of
eqs. (4.5--4.6),  with the
upper sign, exist only for  the choice of
$C=0$. Such solutions have the  property
$\P_\infty\ne\infty$ and  as $\P\rightarrow \P_\infty$,
  $U(\P)$ approaches zero
 faster than $   (\P-\P_\infty)^2$.  We shall see that solutions of this type
correspond to    the case   of  `dilatonic'-type electric solutions  (Section
4.2)
and a
class of `moduli'-type  magnetic solutions (Section 4.3).

Solutions with the lower sign in  eqs.(4.5--4.6) exist for
$C=0$ or $C\ne 0$, depending  on the
nature of  $U$. If $C=0$ one obtains
$\P_\infty= \infty$ and as  $\P\rightarrow \infty$,
 $U\rightarrow 0 $ faster than  $\P^{-2}$ .
 We shall see that such properties are found for a class of
`moduli'-type electric solutions and  a special case of the `dilatonic'-type
 magnetic solution.
On the other hand, there also exist  regular, finite energy solutions
for  a specific value  of $C=C_0> 0$; this is the case  if   $\P_\infty=\infty$
and  as $\P\rightarrow\infty\ $,
  $U(\P)-{1\ov2}C_0$  approaches  zero faster than $\P^{-1}$.  %\foot{Note that
%the case with $C\ne 0$ is special, since it allows for non-positive $U(\P)$
%which should,  however,  be bounded  from below by a  finite negative constant
%$-{C\ov 2}$.}
This case will turn out to correspond to
 an instructive example  of a `dilatonic'--type  magnetic solution.

\subsec{ `Dilatonic' solutions}

Let us consider several  particular examples of $f$. In the following we shall
express  explicit solutions in terms of the radial coordinate
$r$ (rather than in terms of
$x={1/  r}$).
The simplest example is the  one
of the tree level dilaton  coupling
$$  f = \e{-2\P} \ , \ \ \ \ \ \P=\p   \ . \
\eq{4.8}  $$
The solution is given by
 eq. (4.5) with  the upper sign  and   $C=0$ (in order to avoid a singularity
at finite $x$). The explicit form of the solution, its  mass $M$,   charge $Q$
 and the
scalar charge $D$ (see eqs.(4.6) and (4.7)) are  then:
$$\p= \p_0 - \ln ( 1 +  {M\ov r})  \  , \  \ \ \
M\equiv \E=  |q| {\rm e}^{\p_0},\eq{4.9} $$
$$ \  E(r) = { Q \ov (r + M )^2 } \ \ , \  Q= M{\rm e}^{\p_0} \  , \ \
D = - M \ .  \eq{4.10} $$   Note that the electric field and the
 effective  string coupling
$e^\p$  are  regular everywhere. The string coupling
grows from zero at $r=0$ to a finite value ${\rm e}^{\p_0}$ at large distances.
The small distance region is thus a
 {\it  weak coupling} region. Therefore it is consistent to ignore
 the non-perturbative potential $V$ in this region.
We shall see in Section 6 that once the potential $V$
is included, $\p$ will be evolving to its minimum at large $r$.
The   solution  (4.9--4.10) is  stable  since the condition  (3.12)
 is satisfied for (4.8).

The  regular  magnetic solution with the finite energy, a counterpart
of the electric solution
(4.9), is obtained  from the lower sign solution of eq. (4.5)  where
 $f$ in (4.8) is replaced  with   $f^{-1} $, and $q$ with $h$, and
  $C=0$. Then
$$\p=\p_0+\ln (1+{M\over r})\ ,\ \  \E\equiv M=|h|{\rm e}^{-\p_0}\ ,\ \  D=M \
. \ \
\eq{4.11} $$
  The fact that in the magnetic case we got a regular, finite energy solution
 with the same choice of $C$ as in the electric case
  is exceptional;  it is a  consequence of
 the property that for  $f$ in (4.8) $\ f  \ra  f^{-1}
$   corresponds to $\p \ra  - \p$.
 The small distance  region  of the magnetic solution (4.11) is a
strong coupling region
and  hence  there
non-perturbative corrections (inducing  $V\neq 0$) can be significant.
 Like the corresponding regular electric solution, this
magnetic  solution is also stable.\foot{Similar  abelian magnetic solution  for
$f= \e{-2\p} $  considered in  \bizon\ was embedded in a class of
non-abelian solutions and was found to be unstable.  The condition of stability
in the
non-abelian case \maison\bizon\  is different from (3.12)
 because of the additional  non-abelian  term in the
potential $ \cal V$ in (3.10). }
%%%%%%%%%%%%%%%%%%%%%%%%%%%%%%%%%%%%%%

The expressions for the dilaton in the electric
 (4.9--4.10) and  the magnetic (4.11) solutions
coincide  with the expressions for the dilaton  in the electric and
magnetic black hole solutions of \gibb\ghs\  if  $r$ is identified with the
coordinate $\hr
 =r -  M$, where $r$ is defined outside the horizon.  In terms of $r$
eqs.(4.9) and (4.11) give
asymptotic    large $r$ expressions for the dilaton  of \gibb\ghs.
For example,  in the case of the
electric  black hole  solution  the  metric (in
the  Einstein frame) takes the form (see, {\it e.g.}, \horow)
 $$
 ds^2 = -(1- {m\ov \hr})(1 + {M \ov \hr})^{-2} dt^2 +
(1- {m\ov \hr})^{-1} d\hr^2 + \hr^2 d\Omega^2
\ , $$
while the expressions for the dilaton and the electric field  coincide with
 (4.9) and (4.10)
where $r$ is
replaced by  $\hat r$.
The physical mass is $\m = M + m$ and $Q=(M\m)^{1/2}$.
As was discussed  in \horow\ the small $\hr$ region is a weak coupling
 region for the electric
solution but a strong coupling region for the magnetic one (which is obtained
from the electric
 solution by the duality transformation $\p \ra -\p, \ F_{\m\n} \ra {\rm
e}^{-2\p}  F^*_{\m\n}
$).

%%%%%%%%%%%%%%%%%%%%%%%%%%%%%%%%%%%%%%%%%%%%%%%%%%%%%%%%%%%%%%%%%%%
A simple,  but important  generalization of  (4.8)
corresponds to
$$  f = \e{-2\P} +   b \   \ , \  \ \ \ \P=\p \ ,
\eq{4.12}  $$
where the constant $b$  can be interpreted, {\it  e.g.},  as
 a contribution of  threshold corrections to $f$  (see (2.4--2.5))
 in the case when the  space dependence
of the modulus   field can be ignored.   In the case of  toroidal
compactifications and a class  orbifolds with a fixed value of the modulus
field the coefficient
$b$ is   generically   positive  (negative) for the abelian (non-abelian)
 gauge group factors. Assuming first that $b >0$, we get the electric solution
from  eqs. (4.5--4.6)   with the choice of the upper sign and $C=0$:
$$  [-\sqrt { \e{-2\p}  +   b}
+ {\sqrt b} {\rm Arcsinh} ( {\sqrt b}\e{\p} )]^{\p(r)}_{\p_0}  = -{|q|\ov r }
\ ,   \eq{4.13} $$
$$   \E  =
{|q|\ov \sqrt b} {\rm Arcsinh} ( {\sqrt b}\e{\p_0} )
 \ .   \eq{4.14} $$
 Note that  $\p_\infty = -\infty$,
{\it i.e.},  the string coupling
  is increasing with $r$  from zero to a finite value $\e{ \p_0} $
and the energy is finite.
The stability condition (3.12) reduces to
$ { \del^2\ov \del\p^2} f^{-1}\geq 0 $.
It is satisfied  if $\p_0$ is such that  $\e{-2\p_0} -   b \geq 0$.

In the case of          $b< 0$   and the
boundary condition  $\e{-2\p_0} > |b|$,
 eqs. (4.13--4.14)  still apply  with   $b$ replaced by $|b|$ and
${\rm Arcsinh}$  -- by ${\rm Arcsin}$. This  solution  again corresponds
 to the   upper sign of eqs. (4.5--4.6) and $C=0$.
Since we assume that the boundary  value  $\p(r=\infty) = \p_0$ is such that
$f =\e{-2\p} -  |b| $ is always  positive,
we again have $\p_\infty = -\infty$.  This solution
is stable if the maximal value of the string coupling   small
enough, {\it i.e.},
$\e{\p_0} < \sqrt{|b|}$.

Interestingly, for $b<0$ and the boundary  condition  $2\e{-2\p_0} < |b|$,
 a regular, finite energy  solution can be obtained from eqs. (4.5--4.6)
 with the lower sign and $C=-q^2/b>0$. Note, however,
  that in this case the solution is unstable.

In contrast to the case of $f= {\rm e}^{-2\p}$
 (eq.(4.8)), for $f$ in (4.12) one cannot obtain a regular, finite energy
 magnetic solution from a regular electric one by
simply changing the sign of $\p$.
Recall, that the  magnetic solution
 for (4.12)  is given by  eqs. (4.3-4.5)
 with  $f$  replaced by $1/f$, and  $q$ by $h$.
 By simply taking $C=0$  one cannot obtain a finite energy solution. It turns
out  that the only
finite energy, regular  solution exists for  $b<0$ and the choice of $C=-h^2
b>0$. In
this case:
$${1\ov \sqrt{-b}} [{\rm Arcsinh} ( \sqrt{-b}\e{\p} )]^{\p(r)}_{\p_0}
=  {|h|\ov r } \ ,   \eq{4.15} $$
$$\E= |h| (\sqrt { \e{-2\p_0}    -b}-\sqrt{-b})
\ .   \eq{4.16} $$
Note that the energy (4.16) is always lower than the corresponding one
(eq. (4.11)) with $b=0$.
For  $b>0$  one  gets  a  regular solution  with
$\p_\infty = \infty $
 (eqs. (4.5--4.6) with the lower sign  and $C=0$).
However,   now  the
energy is infinite. It turns out that  in this case there is no way of
 adjusting the coefficient $C$
in order to get a finite energy solution.

Finite energy, regular `dilatonic' electric and magnetic
solutions with $f$ in (4.12) are plotted in Figure 2a and Figure 2b,
respectively.
%%%%%%%%%%%%%%%%%%%%%%%%%%%%%%%%%%%%%%%%%%%%%%%%%%
\subsec { `Modulus' solutions}
Our next example is
$$ f(\P) = p^2( \ch  a\P +  s )^2 \ . \eq{4.17} $$
 For large negative $\P$ and $a=1$  this function  is the same as in (4.8).
 Being  symmetric under the `duality' symmetry   $\P \ra - \P$
this $f$ (with  $\P = \vp$, $a= 1/\sqrt 3$)  models well
  the
modular invariant  coupling function $f_2(\vp)$
in (2.5).\foot{ In  the case if  string theory is invariant under
the  conjectured `dilatonic' $S$-duality \font\ the coupling (4.17)
(with $a=1,\ s=0$) may serve
also as a model of a non-perturbative duality invariant modification
of the tree-level dilaton coupling function
(4.8).}
 Note, that here $f_2(\vp)=f_2(-\vp)$,
  $\ f_2(\vp) \ra -{1\ov 3} \pi b_0{\rm e}^{\pm2\vp/\sqrt{3}}
 $ for $\vp \ra \pm \infty $  and  $b_0$
is generically   negative (positive)  in the
case of the abelian (non-abelian) gauge group.
A non-zero  constant $s$ in (4.17) may  be
considered as accounting for  a modulus independent contribution to  $f$.

 Without loss of generality one can  fix the boundary condition
$\vp_0\equiv \vp(1/r=0)>0$.
Namely, due to the duality symmetry $\vp\to-\vp$  solutions
with  the boundary condition
 $\vp_0<0$ are related to the ones with $\vp_0>0$. %\foot{Note, however, that
%%the
% solutions with $\vp_0<0$ obtained by duality symmetry have the energy larger
% than the solutions with $\vp_0>0$
%  displayed in the text. Thus, the solutions presented below
%are the minimal energy solutions.}
 Then  for the regular, positive energy
 electric solutions   one finds from eqs.  (4.5--4.6)  (with the
 {\it lower} sign and
$C=0$)
$$[ \sh a\vp  +  sa\vp]^{\vp(r)}_{\vp_0} =   {|q|a\ov |p|r}    \
 \ ,   \eq{4.18} $$
$$\E
={ 2|q|\ov a|p|\sqrt{1-s^2}} [{\rm Arctan }({\sqrt{1-s^2}\ov 1+s }
\th {a\vp\ov 2}) ]^{\vp_\infty}_{\vp_0} \  , \ \ \ s^2<1   \ ,
% ={ 2|q|\ov a|p|\sqrt{s^2-1}} [{\rm Artanh }({\sqrt{s^2-1}\ov |1+s|}
%\th {a\vp\ov 2}) ]^{\vp_\infty}_{\vp_0} \ \ , \ s^2>1}   \ ,
\eq{4.19}
$$
$$ E=  { q\ov  r^2 p^2[ \ch  a\vp (r)  +  s ]^2  } \ ,\ \
 Q=  {q\ov p^2( \ch  a\vp_0 +  s )^2}\ ,  \ \
  D =
{q\ov  |p|( \ch  a\vp_0 +  s )} \ . \eq{4.20}  $$
For $s^2>1$ the expression for the energy is the same as the one in eq. (4.19),
however, $\rm Arctan$ is replaced by $\rm Arctanh$ and  $1-s^2$ by
$s^2-1$.  For  $s>-1$ the solution is regular for {\it any} $\vp_0>0$.
On the other hand, for $s<-1$,  the  regular solution  exists when  $\vp_0>0$
satisfies $ \ch  a\vp_0 +  s  > 0 $.
 The condition of stability (3.12), {\it i.e.},
$ {d^2 \ov d \vp^2} \ {f }^{-1}   \geq 0  $,
is satisfied if  $\vp_0$ is such that
$ 2 {\rm cosh}^2 a\vp\  - \ s\ \ch a\vp \ - 3\  \geq 0  $
for all values of $ \vp (r)  $.
Since  there always exists a choice of $\vp_0$ for which both
of the above constraints  are  satisfied  we get a class of stable, regular
finite energy electric  solutions.

In   general, the electric solution  (4.18--4.20) has
 the property that it
increases from a positive finite  value $\vp_0\equiv \vp (1/r=0)>0$ to
$\vp_\infty\equiv\vp(1/r=\infty)=+\infty$. Namely, in the core of the
solution, {\it i.e.}, as $r\rightarrow 0$, a
 decompactification ($\vp\rightarrow
\infty$)
takes place.  We would like to draw an analogy with the corresponding
`dilatonic' electric
solutions with $f$ in (4.12):  the role of
the weak coupling  at the core of  the `dilatonic' electric solution  is
now   played  by the
decompactification  at the core of the `modulus' electric solution.\foot{
For a special case with
    $s=0$   we have  $ f = p^2 {\rm cosh}^2  a \vp$
and eqs.
(4.18) and (4.19) take    a more explicit form
$$\vp  = {1\ov a} {\rm Arcsinh} (\sh a\vp_0   +{|q|a\ov |p|r })\     ,  \ \ \
 \E
= { 2|q|\ov a|p|} [{\pi \ov 4 }-{\rm Arctan }(\th {a\vp_0\ov 2})   ]\ . $$
The boundary condition $\vp_0>0$ is chosen. The solution is stable for
$\cosh a\vp_0 \geq \sqrt {3\ov 2}$.}

We have demonstrated that  stable, regular,  finite energy
electric solutions  exist  for the  duality invariant
function $f(\vp)$ in (4.17).
On the other hand, it turns out that there are  no regular, finite
 energy magnetic solutions
corresponding to  such $f(\vp)$, unless  $s =-1$. In the latter case:$$
f=4p^2{\rm sinh}^4 {a\vp\ov 2} \ . \eq{4.21} $$
It is crucial, that now  $f(0)=0$ and the point $\vp=0$  corresponds to the
minimum of $f$.  Here one gets stable, regular,
finite energy solutions for  {\it  both }  the  electric {\it  and} the
magnetic cases.
The explicit form of the electric solution
(lower sign and $C=0$ in eqs.
 (4.5--4.6))  is:
$$(\sh a\vp  - a \vp) -(\sh a\vp_0  - a \vp_0 ) ={{|q|a\ov |p|r}} \  \ ,
\eq{4.22} $$
$$ \E ={|q|\ov a|p|} (\coth {a\vp_0\ov 2} -1 ) \   . \eq{4.23} $$
 Without loss of generality  we can choose the
 boundary condition $\vp_0>0$ and the
 solution
 has  again the property $\vp_\infty\equiv \vp(r= 0 )=+\infty$, {\it i.e.},
  as $r\rightarrow 0$, a decompactification ($\vp\rightarrow \infty$)
  takes place.
 The solution with  the boundary condition
 $\vp_0<0$  is  related to the one of eqs.
(4.22--4.23) by the duality symmetry
 $\vp\rightarrow -\vp$.

The regular  positive energy magnetic solution is obtained from
(4.5--4.6) with the {\it upper} sign and $C=0$:
$$  \coth {a\vp(r) \ov 2} -\coth {a\vp_0 \ov2} =  {|p||h|a\ov r} \ , \ \ \
  \eq{4.24} $$
$$ \E= {|h||p|\ov a} (\sh a\vp_0  - a \vp_0 ) \ \ . \eq{4.25} $$
We chose again
 the boundary condition $\vp_0>0$.  Now,  $\vp_\infty
=0$, {\it i.e.}, the magnetic solution corresponds
 to the compactification at the self-dual point $\vp=0$.
 Interestingly, for the duality invariant $f$  with $f(0)=0$ ({\it e.g.},
 given by (4.21)) the electric (4.22--4.23) and the magnetic
(4.24--4.25) solutions have  complementary features, similar to the ones of the
dilatonic solution with $f$ in (4.8). Now,  however, the role   of the
strong-weak coupling  regions is
played by the compactification - decompactification regions.\foot{
Analogous stable, regular,   finite energy  electric and magnetic solutions
exist  also  for another example of duality invariant $f$ with the property
  $f(0)=0$, namely
$  f = p^2 {\rm sinh}^2 {a \vp } \ .$
 In this case the electric solution (lower sign and $C=0$ in eqs. (4.5--4.6))
 is given by
$$ \cosh (a\vp)-\cosh (a\vp_0)={|q|a\ov |p|r}\ , \ \
\E = - {|q|
\ov |p|a} \ln \tanh ({a \vp_0 \ov 2})  \ ,$$
while the magnetic solution (upper sign and $C=0$ in eqs. (4.5--4.6))
 is
$$ {\rm ln } [ {\tanh ({a \vp\ov 2}) \ov \tanh ({a\vp_0 \ov 2}) }] =
 - {|p| |h|a\ov r} \  , \ \  \E= {|h| |p| \ov a } ( \cosh a\vp_0 - 1 ) \ .
  $$
The qualitative behaviour of the above solutions
is the same as of  (4.22--4.25). Namely,
 as $r\rightarrow 0$ the
electric and magnetic solutions correspond to decompactification
($\vp\rightarrow \infty$)
and  compactification
at the self-dual point ($\vp\rightarrow 0$), respectively.}

One can convince oneself that the finite energy electric {\it  and}
 magnetic  solutions with
 qualitatively  the same behaviour  exist
 for a general positive definite, duality invariant function
$f$ with the following properties: $f(\vp)$ has the minimum at $\vp=0$,
$f(0)=0$, and as $\vp\rightarrow\infty$, $f$ grows faster than $\vp^2$.
 In Figure 3 we show the  explicit numerical
solutions for  the `realistic' example  of $f$   corresponding to the toroidal
and a
class of orbifold compactifications (see eqs. (2.4--2.5))
$$f= b_0 \ln  \left[{ (T + T^*) |\eta (T)|^4\ov 2 |\eta (1)|^4}\right]\ \ \ ,
\ \ b_0={\cal O} (1/100),\ \ \ T=\e{2\vp/\sqrt3} .
\eq{4.26} $$

%%%%%%%%%%%%%%%%%%%%%%%%%%%%%%%%%%%%%%%%%%%%%%%%%%%%%%%%%%%%%%%%%
\newsec { Case of zero scalar potential: solutions with two  non-trivial scalar
fields}

\subsec { General remarks  and magnetic solution}

In this section we   shall study  the solutions in  a more `realistic'  case
when both  the dilaton and the modulus field can change in space.
We shall choose  the  coupling function in the following  form
$$  f(\p,\vp) =  f_1(\p) + f_2 (\vp)  \ , \eq{5.1} $$
which is  a generalization of  the  perturbative expressions (2.4--2.5)
where
$f_1= \e{-2\p} $  and  $f_2=
b_0 \ln [(T+T^*) |\eta(T)|^4]+b_1$, with  $T= \e{2\vp/\sqrt3}
$.
The  presence of a non-perturbative potential  will be ignored.
The system of equations  for the two scalars $\P_i=(\p,\vp) $  in the
case of the electric solution takes the form  (eq. (3.6) with the
 $ U=-{1\ov 2}q^2(f_1+f_2)^{-1}$ and  $V=0$):
$$  \P_i''   + {q^2 \ov 2 (f_1+f_2)^2} {d f_i\ov d\P_i} = 0   \ ,  \ \ \ \
i=1,2\ .  \eq{5.2} $$
The first integral of this system
$$
%{1\ov 2}\P_1'^2+{1\ov 2}\P_2'^2+U=
{1\ov 2}\P_1'^2+{1\ov 2}\P_2'^2-{q^2\ov {2(f_1+f_2)}}=
C=\hbox{const.}\ \  \eq{5.3} $$
remains quite complicated.

The system of eqs.~(5.2)
reduces to the one-scalar case  considered in the
previous section  if   one of the scalars  $\P_i$ is
fixed  to be at the extremum of the corresponding  function $f_i$.
 While  the  tree-level dilatonic
coupling $f_1$ in (5.1) does not have a local extremum
(and thus the dilaton cannot be `frozen'
at a constant value) the  modulus coupling  $f_2$
does have  an extremum  at $\vp=0$.
If   $\vp=0$ then (5.2) reduces to the  case of the dilatonic coupling
(4.12) discussed in Section 4.2.

In the next subsection we shall  find   electric solutions  of eq.
 (5.2)  using a  perturbative approach, {\it i.e.}, by
assuming  that $f_1 \gg f_2$. This assumption is satisfied, in fact,
 in the case
of the threshold correction in  (2.5).
We shall see that in  such a case one can  reduce the
system  of the two  second-order coupled differential equations (5.2) to a
set of { \it two first-order} coupled differential equations.

Let us first consider, however,  a simpler case of
   magnetic solutions.
The corresponding  equations (3.6)  for the magnetic case  with $f$ given
by eq. (5.1) take the  form of the two {\it decoupled }
second order differential equations:
$$  \P_i''   - \ha {h^2  }  {d f_i\ov d\P_i} = 0   \ , \ \  \ \ i=1,2  \ .
 \eq{5.4} $$
Since the total energy (3.8)  for the magnetic case is now the sum of
 separate  contributions  for  each $\P_i$,
it is minimized  if  one combines the minimal energy
 magnetic solutions   for each
$\P_i$ independently.  Such solutions were already discussed  in Section 4.
Let us represent $f$ in
 (5.1)
in the form
 $$ f= {\tilde f}_1 + {\tilde f}_2 \ , \ \ \ {\tilde f}_1=  \e{-2\p} + f_2(0) \
,
\ \ \   {\tilde f}_2  = f_2(\vp) - f_2(0)  \  \ . \eq{5.5} $$
With this choice the duality invariant function $\tilde
f_2(\vp)=\tilde f_2(-\vp)$
has the minimum at $\vp=0$ with
$\tilde f_2(0)=0$ and thus
corresponds to the finite energy,   magnetic `modulus'
 solutions as discussed
in Section
4.3.  For example,
for the `realistic' choice (2.4--2.5),
 $\ \ \tilde f_2(\vp) =  \ b_0\ln[(T+T^*)|\eta(T)|^4/(2\eta(1)^4]$  (eq.
(4.26)),
the corresponding  solution is depicted in Figure 3 (dashed line).
On the other hand,  we saw in Section 4.2 that in the case of
${\tilde f}_1$  (eq. (4.12) with $b=f_2(0)\not=0$) the `dilatonic' magnetic
 solution for $\p$  has
a regular, finite energy solution (4.15--4.16)  (Figure 2b) only when
$b=f_2(0) < 0$. For the case of toroidal compactifications
 and a class of orbifold compactifications (with $f$ defined in eqs.
 (2.4--2.5)) the constraints on $\tilde f_1$ and $\tilde f_2$
can be satisfied  for an abelian gauge group ($b_0<0$ in eq. (2.5)) if
the modulus-independent  threshold correction $b_1$ in
 (2.5) is such that $f_2(0)<0$. Note,
however, that although   not implausible,  the constraints for existence of a
finite energy
   `dilatonic' magnetic solution might be difficult to satisfy.

\subsec { Electric solutions: perturbative  approach }
To find electric solutions with  both fields being  non-trivial
fields it turns out to be useful
to draw an analogy with the scalar field solutions in  the supersymmetric
theories.  Namely, due to the existence of the first integral (5.3)
one can represent the energy of the system in a simplified form assuming
there exists a function $W(\p,\vp)$ (`superpotential') which is related to the
potential
$U=-{1\ov 2}q^2(f_1+f_2)^{-1}$ in the following way:
$$2C-2U=\left({{\del W}\over {\del \p}}\right)^2
+\left({{\del W}\over {\del \vp}}\right)^2\ \ \ .\eq{5.6} $$
In this case the energy of the system can be rewritten as:
$$\E={1\over 2}\int_0^\infty dx\left[(\p'\pm{{\del W}\over {\del \p}})^2+
(\vp'\pm{{\del W}\over {\del \vp}})^2\right]\mp
(W_{x=\infty}-W_{x=0})-\int_0^\infty dx\  C\ ,
\eq{5.7} $$
 where $W_{x=\infty}$ and $W_{x=0}$ denote the values of the superpotential at
  $\p (x=\infty),\ \vp(x=\infty)$ and  $\p (x=0),\ \vp(x=0)$, respectively.

The energy (5.7) is extremized if:
  $$\p'=\mp{{\del W}\over {\del \p}}\ \ \ ,\ \
 \vp'=\mp{{\del W}\over {\del \vp}}\ ; \eq{5.8} $$
then
$$\E=\mp [W_{x=\infty}-W_{x=0}] -\int_0^\infty dx \ C\ .  \eq{5.9} $$
 Thus, one has reduced   the system of two coupled second order   differential
 equations (5.2) to  a more tractable system of
 two  coupled first order differential equations (5.8).
 Equations
of motion (5.8)  are analogs of the Bogomol'nyi equations and
eq. (5.9) is an analog of the Bogomol'nyi bound.\foot{Note  a clear
similarity with supersymmetric configurations, {\it e.g.}, global
supersymmetric domain walls \cqr .}

 Such a  significant simplification can
  take place only if it is possible
 to find a superpotential $W$ satisfying eq. (5.6).
We will now  determine the  explicit form of  $W$  using the approximation
$f_1\gg f_2 $.\foot{This will turn out to be the case for
(2.4--2.5)  as long as the boundary condition  $|f_2(\vp_0)|\e{2\p_0} \ll 1$ is
satisfied.}
Expanding $W$ in powers of $f_2(\vp)$,  $\ W=W_0(\p)+W_1(\p,\vp)+\ldots\ , $
it is easy to show that $$ W(\p, \vp)=\int
_{\p_a}
^\p {d\p}\sqrt{{q^2\ov f_1(\p)}+2C}-{1\ov 2}{q^2{f_2(\vp )}}
\int
_{\p_b}
^\p {{d\p}\ov {f_1^2(\p)\sqrt{{q^2\ov {f_1(\p)}}+2C}}} +\ldots\ .
\eq{5.10} $$
The equations of motion (5.8) then take  the form
 $$\eqalign{\p'&=\mp\sqrt{{q^2\over{f_1(\p)}}+2C}\,+\ldots\,,\cr
\vp'&=\pm q^2{{\del f_2(\vp)}\over{\del \vp}}
\int
_{\p_b}
^\p {{d\p}\ov {f_1^2(\p)\sqrt{{q^2\ov {f_1(\p)}}+2C}}}
+\ldots\ \ \ .}   \eq{5.11} $$
Note that the lower boundary value $\p_a$ in the first integral in eq. (5.10)
 is arbitrary
and the equations  of motion (5.11) as well as the energy (5.9) do not depend
on it.
However, the solution does depend on the
lower boundary value
 $\p_b$ in the second integral in eq. (5.10). It should
be chosen  so that  to minimize the energy (5.9) and yield a regular solution.
Thus, we have traded the two of the  four boundary conditions on  ($\p ,\
\vp$) for the two  constants $C$ and $\p_b$.

Let us   illustrate the above approach
for  $f$ defined in eq. (5.1)  with $f_1=\e{-2\p} $ and $f_2 $  such
%(see eqs. (2.4--2.5))  and using the
that the approximation $f_1\gg f_2$ is valid.
 In this case  (5.10)  gives
$$W(\p ,\vp )= |q| (\e{\p} -\e{\p_a} ) -{1\ov 6}|q|f_2(\vp)
(\e{3\p } -\e{3\p_b} )\ \ \ \  .
\eq{5.12} $$
While the first equation of motion (5.11)
(upper sign and $C=0$) corresponds to the standard
`dilatonic' electric solution (4.8)
with the property $\p_\infty=-\infty$,
the second equation in (5.10) takes the form:
$$\vp'={1\ov 6}|q|{\del f_2(\vp)\ov {\del \vp }}( \e{3\p} -\e{3\p_b} )\ \ \ \ .
\eq{5.13} $$
In the following we choose $\p_b=-\infty$ which turns out to  correspond to a
regular solution with the energy:
$$\E=|q|\e{\p_0} \left(1-{1\ov 6}{f_2(\vp_0)}\e{2\p_0} \right)\ \ \ .
\eq{5.14} $$
One can then rewrite
 eq. (5.13) as:
$$(\vp')^2={1\ov 6}|q|\e{3\p} {d  f_2\ov d x}  \ ,\ \eq{5.15} $$
 where ${d f_2\ov d x}={\del f_2(\vp)\ov\del\vp } \vp'  \ .$

  Since  the left hand side  of eq. (5.15)
is positive, in order for the right hand side to be
 positive,   ${df_2\ov dx}$ has to be positive  everywhere.
In  the abelian case (see eq. (2.5)),  $f_2> 0$ and  it  grows with $\vp$.
The constraint ${df_2\ov dx}>0$ implies that $\vp$ grows  as
 $x=1/r\rightarrow
\infty$. In other words,
 near the core, the solution tends toward  a larger compactification radius.
On the other hand, in  the non-abelian case, {\it i.e.}, for abelian electric
solutions embedded in the non-abelian gauge sector,
 $f_2<0$ and decreases as $\vp$
increases.
In this case  ${df_2\ov d x}>0$
implies that as  $x\rightarrow \infty$, $\ \vp$ approaches
smaller values. Namely,  near the  core  the solution
tends toward a smaller compactification radius.

If the same  constant  $\tilde f= {\cal O} \left[f_2(\vp_0)\right]$  is added
 to  $f_1=\e{-2\p} $  and
subtracted from function
$f_2$ ({\it c.f.} eq. (5.5)), the solution
should remain the same.  This is indeed the case.
% up to the  order in  perturbation expansion we are considering.
 Note,
that the solution  (5.11) with $\tilde f_1=\e{-2\p} + \tilde f$ yields the
`dilatonic' solution (4.13) with $\p_\infty =-\infty$  and the  energy  (4.14)
 which,  in the limit $\e{-2\p} \gg |\tilde b|$,  is given by
$\E_0=|q|\e{\p_0}
(1-{1\ov 6}\tilde f \e{2\p_0} )$. On the other hand, a solution for $\vp$
depends on ${\del \tilde f_2\ov \del \vp}$ (see eq. (5.15)), and is thus
independent of an additional  constant in
 $\tilde f_2=f_2(\vp)-\tilde f$.   This part of the solution gives the
following
 contribution to
the energy:  $\E_1=-{1\ov 6} |q|\e{3\p_0}
(f_2(\vp_0)-\tilde f)  $. The total energy $\E=\E_0+\E_1$ is therefore
unchanged  (see eq. (5.14)).

In order to find  an explicit expression for $\vp$ let us first choose a
simple   duality invariant function  $f_2=p^2\sinh^2a\vp$ which
 models well (up to an irrelevant constant)
the duality invariant threshold correction  $f_2$ in (2.5).
 We will allow  $p^2$ to be negative as well (this will model the  case of
a non-abelian
embedding).   Using the explicit solution for $\p$ in eq. (5.15) one  finds
$$ \ln [{\th a\vp (r)  \ov \th a\vp_0} ] = {1\ov 6} a^2  p^2 \e{2\p_0}  \left[
 1- {1\ov (1+ {|q|\e{\p_0}
\ov r} )^2} \right] \
.  \eq{5.16} $$
 Clearly, as $r\rightarrow 0$, $\vp$  increases (decreases) for  $p^2>0$ (
$p^2<0$).

In order to confirm this  general behaviour for  duality invariant functions
we  have also evaluated numerically the  solution for $\vp$  with  $f_2$
in (4.26), {\it i.e.}, in the `realistic' example corresponding to toroidal and
a
class of orbifold compactifications.
%Then eq. (5.13) takes the form:
%$$\vp'={1\ov 6}
In this case the results    are given  in Figure 4.

We  shall  now show that
the  above   solutions  with $f_2 >0$  (abelian case) are  unstable,
 while those with  $f_2<0$ (embedding in a non-abelian gauge group) are stable.
This   instability is  a generic property of
the  electric two-scalar
solutions with {\it  positive}  and convex  function $f_2(\vp)$.   In fact,
for
an arbitrary $\vp$  the matrix of second derivatives of $f^{-1}$  in (3.11)
 is of the form $$  \del_i\del_j [   \e{-2\p} + f_2(\vp) ]^{-1} =
$$ $$   {1\ov (\e{-2\p} + f_2 )^3 } \left(\matrix { 4\e{-2\p}
(\e{-2\p} -f_2 )                 &   -4f_2' \e{-2\p}
   \cr    -  4f_2' \e{-2\p}     &
 f_2^3 ({1/ f_2})'' - f_2''\e{-2\p}
\cr}\right)   \ ,
\eq{5.17}
$$
 where $f_2'= {\del f_2 \ov \del \vp}$ and $f_2''={\del^2 f_2 \ov \del \vp^2}$.
For  $f_2 > 0$ and $ f_2'' > 0$   there is a region where
$(f_2^{-1})'' < 0 $  and  (5.17)
is not positive  definite.
This is obvious, in particular,  if $f_2$ is small compared  to $\e{-2\p} $.
 This implies that perturbative  electric solutions corresponding to
the case of the abelian group ($  f_2(\vp)> 0$)  are unstable.
 Since the tree-level dilaton
solution is stable, it is clear that the instability  in this case
 is due to   the modulus sector.

 Note, however, that  the abelian electric solutions
 corresponding to the case of embedding into a  non-abelian
gauge group, {\it i.e.},  to the case of  $f_2(\vp)<0$,
 {\it are   stable}. In fact, in this case
$$ f_2^3 ({f_2^{-1}})'' - f_2''\e{-2\p}  = |f_2''| \e{-2\p}  + f_2^3
(f_2^{-1})''
\ $$ is
positive  if $\e{-2\p}  \gg |f_2|$.

The perturbative approach to obtaining  electric  solutions is
valid for a large range of boundary conditions. In particular,
   as   long as $|f_2(\vp_0)|\e{2\p_0} \ll 1$  the  solutions   have the
features
described above. One should,  however,   keep in mind that for  boundary
conditions $\vp_0\gg 0$
the approximation $f_1\gg f_2$ is not valid anymore.
In this case
one can find special solutions if $f_1(\p)$ and $f_2(\vp)$ have similar
functional dependence.  Then  $\p$  and $\vp$  are  related as
well.  For  $\vp \gg 0$ (a large   compactification  radius $T$) and
$f_2$ given in  (2.5) this is indeed the case. In the region  of large
 $\vp $,
$\ f_2$
 can be approximated as
$$ f_2 (\vp) \simeq  \tilde{b} T = \tilde{b} \e{2a\vp} \  ,\ \ \eq{5.18} $$
 where $ a={1/\sqrt{3}}  , \  \tilde{b}=-{\pi\ov3}b_0$ ($b_0$  is the constant
in eq. (2.5)).
For a solution to exist we have to assume  that the constant $\tilde{b}$
is positive, {\it i.e.}, the special solution below exists only in
the abelian case.   Eqs. (5.2) are  then   satisfied if
$\vp$ is related to $\p$ by
 $$ \p = - a \vp + k  \ , \ \ \ \ \  k = - \ha \ln(\tilde{b} a^2) \ , \eq{5.19}
$$
 and $\p$ is the solution  (4.9)  for  the standard  `dilatonic' case  with
 $f_1$  given by   (4.8), {\it i.e.},
 $$\p= \p_0 - \ln ( 1 +  {M\ov r})  \  , \  \  $$
with the rescaled `charge' $q'$
$$  |q'|={|q|a^2\ov 1+ a^2} <|q| %
%{1\ov 10}q
 \ . \eq{5.20} $$
Note, that this  special  solution has the property that
in  the region $r\rightarrow 0$
 the string coupling approaches  zero and the
compactification  radius approaches infinity.%
\foot{The gravitational analog of this special solution  provides a
particular generalization  of  the solution of \gibb\ghs\  to  the case
 with  an  additional
exponential coupling of the modulus to the gauge field term. Such a solution
   was recently discussed in  \cadoni\ .}

Using the same arguments as above one can show that
this particular solution is unstable, as is
the case in general
 for any electric solution with two scalars and    positive and convex
$f_2(\vp)$, {\it i.e.},  the abelian  two-scalar electric solutions.

%%%%%%%%%%%%%%%%%%%%%%%%%%%%%%
%%%%%%%%%%%%%%%%%%%%%%%%%%%%%%%%%%%%%%%%%%%%%%%%%%%%
\newsec{ Solutions  in the case  with   non-perturbative scalar potential}
%%%%%%%%%%%%%%%%%%%%%%%%%%%
Let us now return to the analysis
of the  magnetic and electric  solutions for    the system  (3.6)
 with a non-perturbative scalar potential $V$
included.  Note that in this case the system  of equations (3.6) does not
correspond to a conservative  mechanical system any more,
and thus has no obvious
integrals of motion.

 We shall mainly consider the non-perturbative potential (eq. (2.10))  due to
 gaugino condensation  in the hidden sector of the gauge group.
 In this case $V$ depends on both the dilaton,  $S= \e{ -2\p} +i\alpha$,  and
the
modulus, $T=\e{2\vp/\sqrt 3} +i\beta$.  In the following we shall take
 the imaginary part of $S$  to be constant  and  the
imaginary  part of  $T$ to be  zero.  In general,
 a   non-perturbative potential should vanish
 in the limit of
small string coupling $\e{ \p}  \ra 0$; {\it  e.g.}, for $V$ in eq. (2.10),
  $V  \ra  \exp[-a_0\exp(-2\p)] \ra 0 $. In addition,  $V$ in  (2.10)  has a
minimum  at
$T_0\sim 1.2$ and  $S_0=\e{-2\p_0} +\alpha_0$. As discussed in section 2.2,
the potential is not convex
everywhere;  in addition to  the   minimum,  it  also has saddle points and
local
maxima.

 The  system of equations (3.6) is   of the form  :
 $$ \p'' + {\del U \ov \del
\p} - {1\ov x^4 } {\del V \ov \del \p} = 0   \ ,
 \eq{6.1} $$
$$ \vp'' +{\del U \ov \del \vp} - {1\ov x^4 } {\del V \ov \del \vp} = 0   \ ,
 \eq{6.2} $$
 where  $U=-{1\ov 2} q^2 f^{-1}$ and  $U=-{1\ov 2} h^2 f$ for the electric and
magnetic   cases,
respectively. Here  again $x=1/r$ and $\p'={{d \p}/{d x}}, \ \vp'={{d \vp}/{d
x}}$.
The solutions of eqs. (6.1--6.2) have to satisfy
 the  stability constraints (3.12). We  shall consider a class of  functions
$f=f_1(\p)+
f_2(\vp)$  (eqs. (2.4--2.5)).
We will  find  that for the minimal energy configurations
 both the dilaton {\it
and}  the modulus  will  asymptotically (at $r\ra \infty$) approach  the values
corresponding to
 the minimum of the potential,
{\it i.e.}, the main role of the non-perturbative  potential is to
  fix   the asymptotic  values of  the dilaton and the modulus
 to be  at its  minimum.
% In the core region, {\it i.e.}, $r\ra 0$,  the solutions  will  correspond to
%the ones discussed in  Section 5.

\subsec {`Dilatonic' case}

For the electric solutions in the absence of a   non-perturbative  potential
(Section 5) and a class of  functions $f$ in (2.4--2.5), the modulus field
$\vp$
varies only slightly. From its boundary
value  $\vp=\vp_0$ at $r=\infty \ $  it
increases (abelian case)  or decreases (non-abelian embedding case)  as one
approaches
the core   $r\ra 0$ (see Figure 4).\foot{Recall,  that only the
solution corresponding to the case of embedding in the
non-abelian theory is stable (see Section 5.2).} In the magnetic case,  $\vp$
also varies  slowly  from  $\vp_\infty=0$    at the origin
to the asymptotic value  $\vp_0$  at $r\ra\infty$ (dashed
line in Figure 3).  We saw that in the case without a non-perturbative
potential
(Section 5) the modulus contribution did not affect the evolution of  the
dilaton $\p$
significantly.\foot{From  the discussion in Section 2.2 it  follows
 that  $f$ in (2.4--2.5) and $V$ in (2.10)   are  both extremal  at
$\vp=0$ (the self-dual point $T=1$). Therefore,   the  eqs. (6.1--6.2) admit a
constant
solution $\vp=0$. Note, however, that at $\vp=0$,   $V$  in (2.10) has a
local {\it  maximum} in the $\vp$ direction\ffont\CFILQ . That means
 this  solution is {\it unstable}.
Namely, in the region $r\ra \infty$  the corresponding Laplace operator for the
linearized perturbations (Section
3.2)  has  a  potential (3.11)  which is unbounded from below and thus,
 should have negative eigenvalues.}

In this section we  shall    therefore neglect a radial variation of the
modulus, {\it i.e.}, we shall simplify the   problem by `freezing out' the
modulus
 and study only  eq. (6.1)
 with $$f=\e{-2\p} +b\   ,\ \  \ \ \ \ V= V(\p) \ . \eq{6.3} $$
Keeping in mind that  the constant $b\ll 1$, we can  ignore its presence,
except in the
 strong  coupling region.

For the electric solution ($U=-{1\ov 2} q^2f^{-1}$) without
 the non-perturbative potential,  the small $r$  region corresponds
 to a weak coupling region,  $\p\ra -\infty$  as  $r\ra 0$
(see Figure 2a and eqs. (4.13--4.14)).
 From eq. (6.1) one then  concludes  that in this region $(x\ra \infty)$ one
can neglect the
contribution of the  non-perturbative potential $V$.  This  is also consistent
with the fact that $V$ is expected to
be negligible  in the weak coupling limit.  Thus, the small
radius ($x=1/r\to\infty$) behaviour
corresponds to a solution  satisfying  $\p'=\sqrt{q^2e^{2\p }+c^2} $ and it is
of
the form:
%he same as the one in  eq. (4.13) expanded  in  $r\ra 0$
%  $$  \p (r \ra 0) =  \ln {r\ov |q|}  + n r + \O(r^2)   \  , \ \ \
% n = - {{\sqrt b}\ov |q|}
%{\rm Arcsinh} ( {\sqrt b}\e{\p_0} )  -  {{b}\ov 2|q|} \ \ .\eq{6.4 } $$
$$\eqalign{\p&=-c|q| (x-x_0)+\log{(2c)}+...\ ,  \ \ \ \ \
c\neq 0\ , \cr
\p&=-\log{|q|(x-x_0)}+....\ ,  \ \ \ \  \ \ \  c=0\ , }      \eq{6.4} $$
where, $c$ and $x_0$ are constants determined by the properties of the
potential.
Note, that the special form of the  solution with $c=0$
which yields a finite contribution to the energy of the system can be obtained
only for a special choice of the parameters of the potential.

On the other hand,  the  asymptotic form  of $\p$ at large  distances
($x=1/r\ra
0$)  can be found by expanding
$$ \p=\p_0 + k_{1}x +l_{1}x^2 + m_{1}x^3 + n_{1}x^4  + ... \   \ . \eq{6.5}  $$
 Eq. (6.1) is then satisfied  to the leading order  in $x$  if
$$ k_1=l_1=m_1=0, \ \  n_1\ne 0  \ , \ \  \  \  ({\del V \ov \del\p})_{\p_0} =0
\ ,
\ \ \ \  \eq{6.6} $$
where $n_1$ is determined by
$$ {q^2\ov 2 }({1\ov f^2 } {\del f \ov \del \p})_{\p_0}  -n_1
({\del^2 V \ov \del \p^2})_{\p_0} =0 \ . \eq{6.7} $$
Thus,
$$ \p(r \ra \infty)   = \p_0 + {n_1\ov r^4} +  \O({1\ov r^5}) \ , \eq{6.8} $$
 and the  asymptotic value of $\p$  should correspond to an
extremum of $V$. Moreover,   $f$  in (6.3)
 is positive  and ${\del f \ov \del \p}$
is negative.  Making  the   assumption that $\p$ should be growing
monotonically with $r$,
 {\it i.e.}, $n_1<0$, we conclude from   (6.7) that
 $({\del^2 V \ov \del\p^2})_{\p_0}$ should be positive, {\it  i.e.},
that $\p_0$ should correspond to   a  {\it minimum} of $V$.  This implies that
even though the
potential $V$  may have local maxima on the way from
 the weak coupling region to   a non-perturbative minimum (as it is
actually the case  for the gaugino  condensation  potentials)
the final point  $\p_0$ of  the `evolution'
of $\p$ from $-\infty$  is at the {\it  minimum} of $V$.
  Note a crucial dependence of this conclusion on
the properties of the function $f$.\foot{It is interesting to
compare this  conclusion to the one  in the case of  time-dependent
  cosmological
solutions.  There   the dilaton  $\p$ is changing with
 time and not with space and the  gauge field  is assumed to be constant
so that   there is no $f$- dependent contribution
to  the equation of
motion (see  \tse\ and references therein).
The existence  of  a local maximum in $V$  seems to preclude a
 natural evolution of the dilaton from a weak
 coupling region  where $V=0$ to  a
`non-perturbative' minimum  at late times
 unless one fine tunes
the initial value of the dilaton `velocity' so that it can  pass over the
maximum to be trapped in
the minimum.}

Similar analysis  of eq. (6.1) can be carried out in the magnetic case
($U=-{1\ov
2} h^2f$).
As we have found in the  $V=0$ case,  the small radius    region  ($r\ra 0$)
 corresponds to a strong coupling  region
($\p\ra \infty$)  (see Figure 2b and eqs. (4.15--4.16)).
That means  that the presence of the potential in eq. (6.1)
 cannot be  neglected anymore.  In the strong coupling region
the non-perturbative  potential (2.10) becomes\foot{
Note that this limiting form of $V$ is the same as
in the case  of  the  potential  corresponding to a  `central charge deficit'
$c\ne 0$, {\it
i.e.},
$V=c\e{2\p} $ (eq. (2.3)).  }
 $$  V (\p \ra \infty) \ra  V_0 \e{2\p}  \
  \ . \eq{6.9} $$
Then (6.1) takes the  following form  in the $r \ra 0$ region
$$ \p'' +  {h^2} \e{-2\p}  - {2\ov x^4 }  V_0 \e{2\p} = 0   \ .
 \eq{6.10} $$
This equation  is solved by a particular solution\foot{In the case of the
non-perturbative potential (2.10) $V_0<0$ and then the additional
 constraint $1+8h^2V_0>0$
 has  to be satisfied in order to have a positive value of $k$.
Detailed  properties of  solutions for other forms of the potential in the
strong coupling region will be  be discussed elsewhere \ctii.
}
$$ \p (r\ra 0)=  \ln {k\ov r }  \ \ , \ \ \ \ \  2V_0 k^4 + k^2 = h^2  \ .
\eq{6.11} $$
The solution (6.11) is the  same (up to a constant)  as the asymptotic form  of
 the solution  with
$V=0$  (eq.(4.15)) in the region
  $r\ra 0$, {\it i.e.},
 $\p(r\ra 0)  =  \ln {|h|\ov r }$.

As for  the large $r$ asymptotics, here we shall again assume that the regular
solutions
 satisfy  $\p(r\ra \infty ) = \p_0$.  Then using the expansion (6.5)
we find   the relations similar to (6.6--6.7):
$$ k_1=l_1=m_1=0\ , \ \ n_1\ne 0  \ , \ \  \  \  ({\del V \ov \del\p})_{\p_0}
=0 \
, \ \ \ \ \eq{6.12} $$
$$  {h^2\ov 2 } ({\del f \ov \del\p})_{\p_0}  +   n_1   ({\del^2 V \ov \del
\p^2})_{\p_0} =0 \ ,  \eq{6.13} $$
and\foot{Similar
asymptotic behaviour of the dilaton was  found  in  the massive dilaton case
in \GH\HH. }
$$ \p(r \ra \infty)   = \p_0 + {n_1\ov r^4} +  \O({1\ov r^5}) \ . \eq{6.14} $$
The assumption    that
 $\p$ is decreasing  monotonically  as $r\ra \infty$  implies that now   $n_1$
should  be
positive.
 Since   ${\del f \ov \del  \p}$  is negative,  we  find  from (6.13)  that   $
({\del^2 V
\ov \del\p^2})_{\p_0}$ should  be positive,  so that,   as in the electric case
,  $\p_0$ should
correspond to the   minimum  of $V$.

We thus conclude that
% the structure of the electric and magnetic
%solutions is not  modified qualitatively
%by the presence of the potential $V$. The
the   important role of the potential is to {\it } fix
the  large distance asymptotic value of the dilaton to be  at the  {\it
minimum}  of $V$.
The energy (3.8) of these
 solutions can be  finite  for a specific  type of the potential.
In addition, the value of the potential at the
minimum  should be  zero, {\it i.e.},   $V(\p_0)=0
$.  If $V(\p_0)\not=0$ (which is, unfortunately, a  generic  property of the
gaugino condensation
potentials)  it is easy to see that
the integral in  (3.8) is divergent at $r=\infty$.  This divergence is simply
due
to the presence of a constant  vacuum energy density proportional to the
cosmological constant
$V(\p_0)$ and thus  can be  ignored in the  present context.

 \subsec{Solutions with  both dilaton and modulus}

We will now  study the general case when both the dilaton and the modulus
in (6.1--6.2)  are changing with $r$.
 Let us consider first the electric solutions. At small
 radius $r\ra 0$ the electric solutions correspond to the weak coupling
region, $\p\ra -\infty$, and thus, the potential $V$ term
can be  neglected in both eq. (6.1) {\it and} eq. (6.2). Thus,
  in the  $r\ra 0$ region  the electric  solution can be discussed along the
lines  in Section 5.2.

Let us now   determine  the  large $r$ asymptotic behaviour of
both  $\p$ and $\vp$.   For the electric solution   we  shall again assume that
 in the large distance  region $\p$ and $\vp$ approach
  constant values $\p_0$ and $\vp_0$.  Using the expansion
$$ \P_i=\P_{0i}  + k_{i}x +l_{i}x^2 + m_{i}x^3 + n_{i}x^4  + ... \  , \ \ \
\P_i= (\p ,\vp)
\  , \eq{6.15}  $$
  one finds that eqs. (6.1--6.2) are  satisfied  to the leading order  in $x$
if
$$ k_i=l_i=m_i=0, \ \  n_i\ne 0  \ , \ \  \  \  ({\del V \ov \del\P_i})_{\P_0}
=0 \ ,
\ \   \eq{6.16} $$
$$ {q^2\ov 2 }({1\ov f^2 } {\del f \ov \del \P_i})_{\P_0}  -\sum_{j=1}^2\ n_j
({\del^2 V \ov \del \P_i\del \P_j})_{\P_0} =0  \ , \ \ \ i=(1,2) \ . \eq{6.17}
$$
As a result,
$$ \P_i(r \ra \infty)   = \P_{0 i} + {n_i\ov r^4} +  \O({1\ov r^5}) \ ,
\eq{6.18} $$
 where the  asymptotic value  $\P_{0i}=(\p_0, \vp_0)$   corresponds to an
extremum of $V(\Phi)$. As follows from (2.5),    $f$
 is positive,   ${\del f \ov \del \p} $
is negative and ${\del f \ov \del \vp} $ is positive (negative) for the
abelian  (non-abelian  embedding) case.  For consistency with the behaviour of
the solutions  at
small $r$ (Section 5.2) we are to assume that  as $r$ increases  $\p$ should be
growing and
$\vp$ should be falling (growing) in  the abelian (non-abelian embedding) case,
 namely,
$n_1<0$, and $n_2>0$ ($n_2<0$). These constraints in turn imply that
in the case  when the  mixed second derivative  term $({\del^2 V\ov \del
\p\del\vp})_{\Phi_0}$
 is small,  the matrix $({\del^2 V \ov \del \P_i\del \P_j})_{\P_0}$ is positive
definite, {\it
i.e.},  both  $\p_0$ {\it and} $\vp_0$   correspond to the  {\it minimum } of
$V$.

We will now show  that the potential $V$ in (2.10) has actually
 the  mixed  second derivative equal to zero,
$({\del^2 V\ov \del
\p\del\vp})_{\Phi_0}=0$.
In fact,  $V$  belongs  to  a class  of  potential which  are
determined in terms of a {\it separable} K\" ahler function
 ${\cal G}={\cal G}_1(S,S^\ast)+{\cal G}_2(T,T^\ast)$ (see eqs. (2.6--2.9)).
  For such a potential
 one can  show that  its  mixed  second derivatives  are always proportional to
 $|{\cal G}_S{\cal G}_T|$, where   $ {\cal G}_S$ and  $ {\cal G}_T$
are derivatives  of ${\cal G}$  with respect to $S$ and $T$.
  Using the
form of the $N=1$ supergravity potential  $V$ in eq. (2.7) one obtains:
$$ {\del^2 V\ov \del S\del T}={\cal G}_S{\cal G}_T\left\{ V+\e{{\cal G}}
\left[{\cal
G}_{T^*} ({\cal G}_{TT^*}^{-1})_T+1\right]\right\}\  + \ {\cal G}_S{\cal
G}_{T^*}
\ \e{{\cal G}} \ {\cal G}_{TT^*}^{-1}{\cal  G}_{TT}\ \ . \eq{6.19} $$
 Thus,  if either  ${\cal  G}_S=0$  and/or
${\cal G}_T=0$ (either one  or both  $S$ and $T$ sectors
preserve supersymmetry),  then  the mixed second derivatives are  zero.
 In the case of  the potential
(2.10)  the  supersymmetry  is preserved at the minimum  in  the dilaton
direction.  Therefore\foot{Recall that $\p$ and $\vp$ are related
 to $S$ and $T$  according to eq. (2.2).},    $\left({\del^2
V\ov \del \p\del\vp}\right)_{\P_0}=0$.\foot{Note that  even if this were
  not the case, the  asymptotic value of the dilaton would
still be fixed at  the minimum. Since  $f_1(\p)\gg f_2(\vp)$,  eq. (6.17)
implies that
  $|n_1|\gg|n_2|$.   Then in the first  equation in  (6.17)
the  contribution of the term with the  mixed   second  derivative could  be
neglected.
Thus the
conclusion of Section 6.1
 that $\p_0$ corresponds to the minimum of the potential along the $\p$
direction,  is still  valid. Note, however,
 that if  the term with the  mixed
second derivative is  not negligible in the second equation in (6.17), then
 $\vp_0$ need not correspond to the minimum
of the potential in  the $\vp$ direction.}
 This  implies that the  modulus field $\vp$  should  also be
 asymptotically  fixed  at the minimum of $V$.

Let us  finally consider  the magnetic solutions.
At small  radius, $r\ra 0$, the magnetic  solutions  have  the strong coupling
region, $\p\ra +\infty$.   The potential $V$ term in eq. (6.1), however,
 {does not}  qualitatively modify  the particular dilaton solution in this
region
(see eq.(6.11)). In addition,  the  potential term can be
neglected  in eq. (6.2)  since it
is  proportional to $\e{2\p} x^{-4}\propto r^2$.
 Therefore, as $r\ra 0$,  solutions
for the modulus  can be discussed along the  lines in Section 5.1.
The large radius behaviour can be studied in the same way as
 for the electric solutions with similar  conclusions.
Thus, in both  the  electric and the magnetic,
  two--field cases the presence of the  potential $V$  term
in the equations of motion (6.1--6.2) fixes
the asymptotic values of $\p_0$ and $\vp_0$  to be at the    minimum   of $V$.

%%%%%%%%%%%%%%%%%%%%%%%%%%%%%%%%%%%%%%%%%%%%%%%%%%%%%%%%%%%

%%%%%%%%%%%%%%%%%%%%%%%%%%%%%%%%%%%%%%
\newsec { Concluding remarks }

In this paper we  have  studied   electric and magnetic configurations with
nontrivial dilaton   and  modulus   fields
  that  are  extremas  of  the
   $D=4$ string effective   action  with  loop (higher genus) and
non-perturbative string corrections  included.
 We have  confined our  analysis  to the  case of the
  abelian  charged solutions associated with  an  abelian gauge group as well
as
abelian charged solutions that can be  embedded into   a non-abelian gauge
theory.

We have found regular, stable,  finite energy  solutions  with both the dilaton
$\p$  and
the modulus $\vp$ varying  with the  radius $r$.
In  the electric solution the dilaton $\p$ always approaches the weak
coupling region in the core, $\p\ra -\infty$ as   $r\ra 0$
(Figure 2a).
The  modulus $\vp$  changes  only slightly from its
 asymptotic value $\vp_0$  at $r=\infty$ (see Figure 4).  For  the  solution
in the case of  an abelian
gauge group   $\vp$    increases, {\it i.e.}, the compactification radius
grows as one approaches  the core region ($r\ra 0$). In other words,
in the core region there is a
tendency towards   decompactification. On the other hand, for  the abelian
electric
 solution embedded into
a non-abelian gauge  theory   $\vp$ slightly  decreases   at small scales,
{\it i.e.}, in the core region  there is a tendency towards
compactification.  Only the latter
 solutions turn out to be stable. The main role of the non-perturbative
 potential $V(\p,\vp)$ is to fix the  asymptotic values  of $\p$ and $\vp$  at
$r=\infty$ to
be  at  the {\it  minimum} of $V$.   This happens  in spite of
 the fact that $V$  may
have other extrema,  corresponding to  saddle points or maxima.
These   charged, finite energy, stable solutions  can be interpreted as
interesting  particle-like field
configurations.

 Regular, {finite energy}  magnetic solutions exist only in the abelian case
and for a specific  range  of the moduli-independent part of
the  threshold corrections in the gauge  coupling  function $f$ (see Section
5.1).  Here, in  the core  ($r\ra 0$), the  solution   approaches the
the  strong coupling region, $\p\ra \infty$ (Figure 2b).
  As   $r\ra 0$  the modulus  tends towards
the  region of  compactification  at  the self-dual point, $\vp\ra 0$ (see
Figure 3, dashed line).
The presence  of the non-perturbative potential  again fixes the  large $r$
asymptotic
values  of the two fields $\p_0$ and $\vp_0$  to be at the minimum of the
potential.

We have also  considered  solutions corresponding to the  models in which  $f$
depends only on
$\vp$ and is  invariant under the modulus  duality  symmetry,
$f(\vp)=f(-\vp)$, or what we called
the  `modulus' solutions.  In the   model with   $f(0)=0$
 and $f$ is growing faster than $\vp^2$ we have found  stable, regular, finite
energy
`modulus' solutions for {\it both} the electric and   magnetic cases.
Such  electric and magnetic solutions  (Figure 3)
  exhibit  analogous  complementary behaviour as  the corresponding
`dilatonic' solutions (Figure 2), with
 the weak coupling ($\p\ra
-\infty$) -- strong coupling ($\p\ra \infty$) regions  replaced by the
compactification ($\vp\ra  0$) -- decompactification ($\vp\ra \infty$) regions.

 The threshold  corrections to $f$ and
the non-perturbative contributions to  $V$  respect the modulus duality
symmetry.
That means the  dyonic  modulus solutions
with a  nontrivial  axion field  of the complex modulus field $T$ can be
obtained by  employing the  $SL(2,{\bf Z})$  $T$-duality symmetry of the full
string effective action.

As for the  duality in the dilatonic sector, it is  present in the  tree level
string
effective action  but  appears to be  violated  by the  perturbative  as well
as  non-perturbative
string corrections (unless it is somehow restored  in some particular cases
in the full string theory
as  was suggested  in \font).
At the tree level,  the
electric and  magnetic solutions are related by  the transformation  $\p \ra
-\p$ (Figure 2, $b=0$ case) what  makes it possible to concentrate only on the
pure
electric or pure magnetic
`dilatonic' solutions.  Tree-level  dyonic solutions   with non-trivial axion
of the
complex dilaton field $S$ can be
found by the $SL(2,{\bf R})$  $S$-duality transformations  as in
\shapere\kallosh.
Since the  dilatonic $S$-duality is, however,  broken by both  perturbative and
non-perturbative  string corrections  discused  in this paper,
 one can no longer   relate the electric and magnetic `dilatonic' solutions
(see
Figure 2, $b\ne 0$ case).

In our  study  we  have  assumed the metric to be flat.
This is consistent  provided  the energy  of the solutions is small enough.
It is important of course to  extend   the analysis of this paper to the
gravitational case,
{\it i.e.}, to find the
 the corresponding  black hole - type   solutions  (generalizing those of
\gibb\ghs\GH\HH) of the
string  effective action  with  the
 threshold corrected   coupling $f$ as well as  the  non-perturbative
   potential $V$ included.

Another direction is to generalize  our solutions to a non-abelian case.  The
non-abelian
dilatonic black
hole solutions, which  are the analogs of  the `ordinary' non-abelian   black
hole solutions
\bartnic\  in  the  case with the  three level  dilaton  coupling
$f=\e{-2\phi} $   were recently
discussed  in \bhym.
 These non-abelian solutions  have finite energy, but are
unstable. It may be interesting to  study their modifications   in the presence
of  a
more realistic  coupling  $f$  and potential $V$.

 At  the  tree-level  of toroidally compactified heterotic
string  theory there
  exist   monopole solutions  discussed  in \khuri.   Since the gauge field is
non-trivial the  dilaton is   changing  in space and
  grows  at  small distances. One could try
to   understand  how such solutions  are  changed   by  perturbative
as well as  non-perturbative  corrections to the  string effective action.

It would   be important    to  find   cosmological analogs of the solutions
presented in this paper.  For a recent  discussion  of cosmological solutions
of
the   superstring
effective action  with  the  one-loop generated  modulus
coupling to the $R^2$ term  see \anton.

Another open problem is how to generalize, to all orders in $\a'$, the
solutions
of the string
effective action with perturbative  as well as non-perturbative
corrections included. At the tree level  a  solution, which is   exact
to  all orders in $\a'$,   can, in principle,  be   obtained by  identifying
the corresponding $2d$ conformal field  theory.
However, once perturbative as well as non-perturbative contributions
to  the effective action are taken into
account, the  equivalence between   extrema of
 the string effective action    and $2d$
conformal field theories  is not established  and  is unlikely to exist.

To conclude,    we would like   to emphasize   again  the  generic features of
our
 stable, finite energy electric
solution.\foot{As was discussed above,  a stable, finite energy magnetic
solution  exists only for
a  certain range of  modulus  independent   threshold corrections.}
It is a  particle-like configuration with the weak coupling region  inside the
core.  It  seems  that
 the proper  dilaton  boundary condition  for  analogous  solutions in string
theory
  should be  $\p\ra -\infty$  at  $r\ra 0$, {\it i.e.},  small string
 coupling at small scales.
This  corresponds to  an appealing  `asymptotic freedom' scenario  in which
 both perturbative and non-perturbative  string corrections  are negligible
 in the small distance region. Thus,
in this region  the tree-level string theory applies, supersymmetry  and other
symmetries are unbroken.\foot{In
the cosmological context the small distance region  corresponds to an  early
time era.}
On the other hand, the growth of the dilaton with $r$ implies that  at   large
distances, or in `our world',  the string coupling  becomes
  relatively strong,  so that  non-perturbative corrections can no longer
be ignored.  In this region
supersymmetry is spontaneously broken and   both   the dilaton and the modulus
fields are  stabilized
at the minimum of the non-perturbative potential.

\bigskip
{\bf Acknowledgements}
\noindent
 We would like to acknowledge  useful  discussions with R. Khuri and D.
Wiltshire. The work was supported in part by
the U. S. DOE Grant No.\ DOE-EY-76-C-02-3071 (M.C),
 NATO Research Grant No. 900-700 (M.C.) and
  by SERC (A.A.T.).

%\vfill
%\eject

\listrefs

%\vfill\eject
{\bf Figure Captions}

\bigskip

\baselineskip10pt
{\bf Figure 1}: Duality invariant function $f=b_0\ln [(T+T^*)|\eta(T)|^4]$
(eqs. (2.5),
(4.36)) versus $\vp$ with $T=\exp ({2\vp/\sqrt 3 })$  for $b_0 = -1$.
\bigskip

{\bf Figure 2}: Electric (Figure 2a) and magnetic (Figure 2b) `dilaton'--type
solutions   $\p (r) $  (eqs. (4.13) and (4.15)) for
  $f=\e{-2\p} +b$  (eq. (4.12))  with  different values of
 the parameter $b$ and $\p(r=\infty)= \p_0 =0$. In Figure 2a  the electric
solutions with $b=0$ and
$b=1/100$  are indistinguishable.   Stable, regular,
finite energy  electric solutions exist
 for $b>- \e{-\p_0} $  while the corresponding magnetic  solutions exist only
for $b<0$.
For the boundary condition
$\p_0\ne 0$ the solutions are numerically different but  the qualitative
behaviour is not changed. The  energy of the electric and  magnetic
solutions increase with increasing $\p_0$.

\bigskip

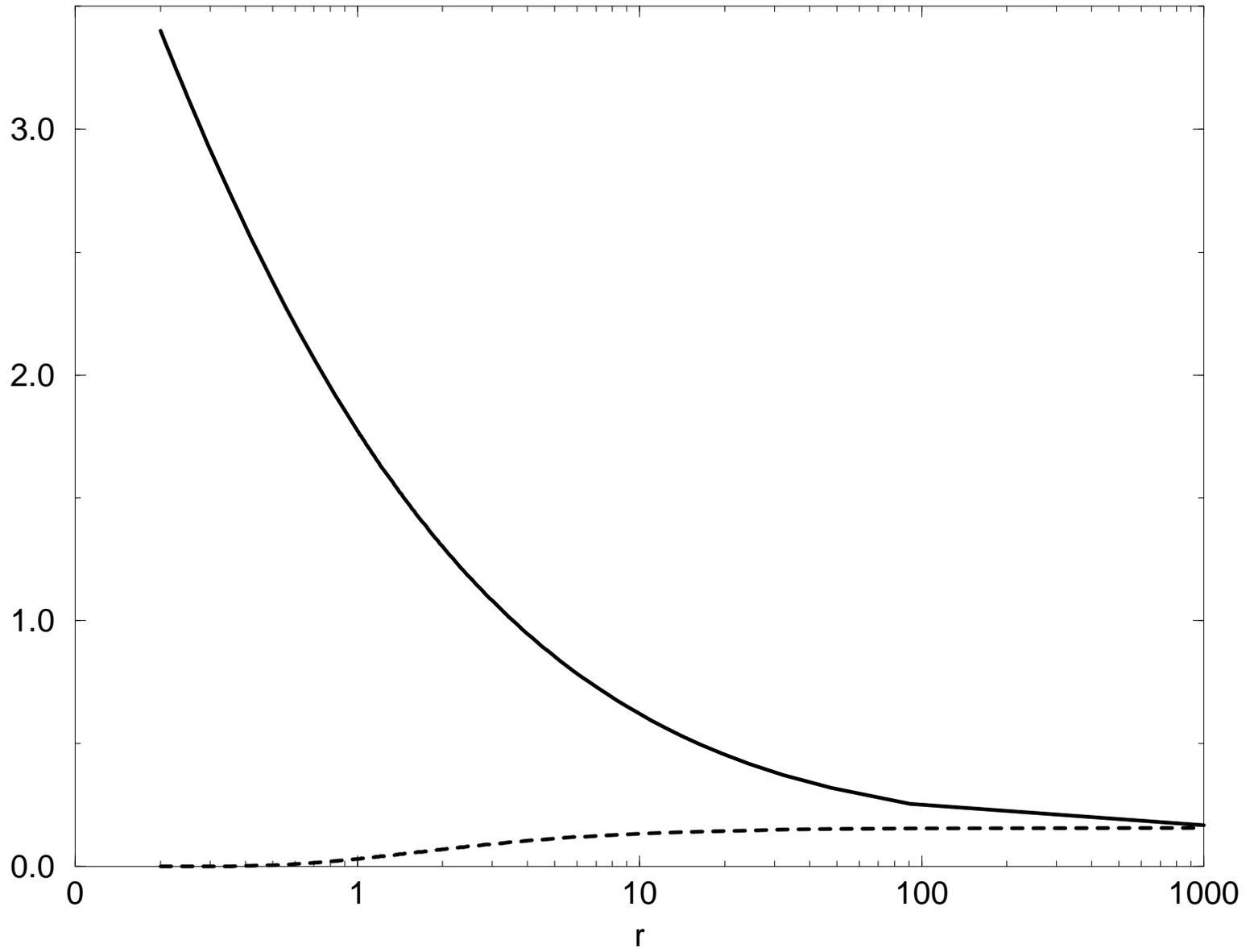
{\bf Figure 3:} Electric (solid line) and magnetic (dashed line) `modulus'-type
solutions for $\vp (r) $  in the case of  $f=b_0\ln
[(T+T^*)|\eta(T)|^4/(2|\eta(1)|^4)]$,  $\ T=
\exp({2\vp/\sqrt 3}) $
(eq. (4.26)). The $r=\infty$ boundary condition  with $T_0=1.2$ or
$\vp_0=0.158$, is  used.  We have chosen  $b_0=-1 $,  the electric
charge $|q|=1$, and  the magnetic charge $|h|=1$. The energy of  the
electric solution is  $\E_0 =3.065$ and  of the magnetic solution
  $\E_0=0.0386$. For
the electric solutions with  $b_0\ne -1$,
$|q|\ne 1$  and the magnetic solutions with $b_0\ne -1$, $|h|\ne 1$ the
 radius $r$ is rescaled by $\sqrt{-b_0}/|q|$ and
$|h|/\sqrt{-b_0}$, respectively,  and the energy -- by  $|q|/\sqrt{-b_0}$ and
$\sqrt{-b_0}/|h|$, respectively. For  the boundary condition
$T_0\ne 1.2$ the solutions are numerically different but the qualitative
behaviour is not changed. For larger  (smaller) $T_0$ the electric energy
decreases (increases),
while the  magnetic one increases (decreases).

\bigskip
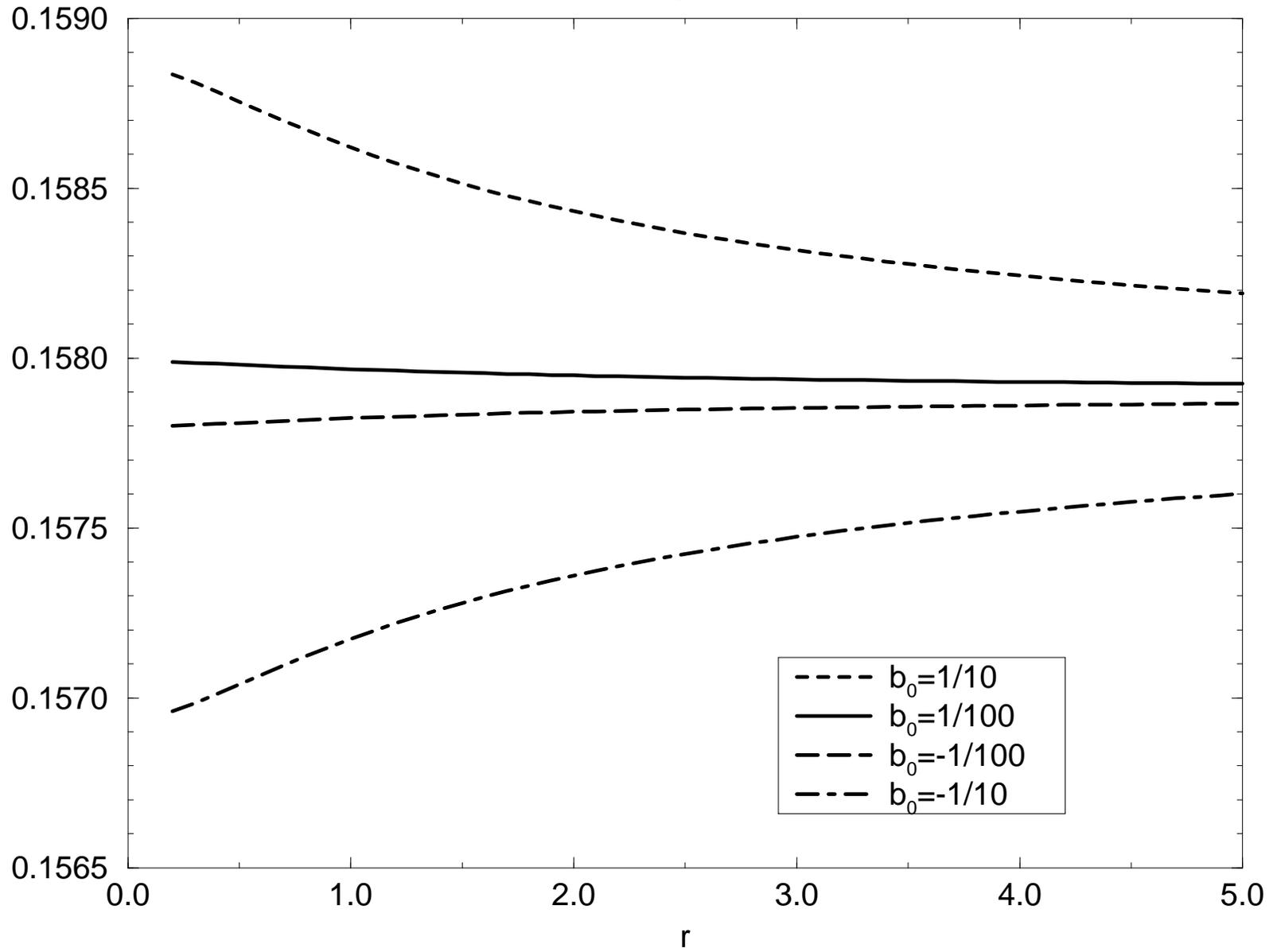
{\bf Figure 4:} The plot of the modulus field  $\vp (r) $
in the case of  the two-scalar  electric solution
with  $f=f_1(\p)+f_2(\vp)$ (5.1),  $\ f_1 =
 \e{-2\p} +b$
(4.12) and $f_2=b_0\ln [(T+T^*)|\eta(T)|^4/(2|\eta(1)|^4)]$
(4.26) with $b={\cal{O}} (b_0)$.
The electric charge is $|q|=1$  and the  $r=\infty$
boundary condition for $\p$ is  $\p_0 =0$. The boundary condition
for $T$ is    $T_0=1.2$ or
 $\vp_0=0.158$.  The  solution for $\p (r) $
is  the same as shown in  Figure 2a.
It turns out (see discussion in Section 5.2) that  only the solutions with
$b_0>0$ (non-abelian  embedding case) are stable.

\vfill\eject
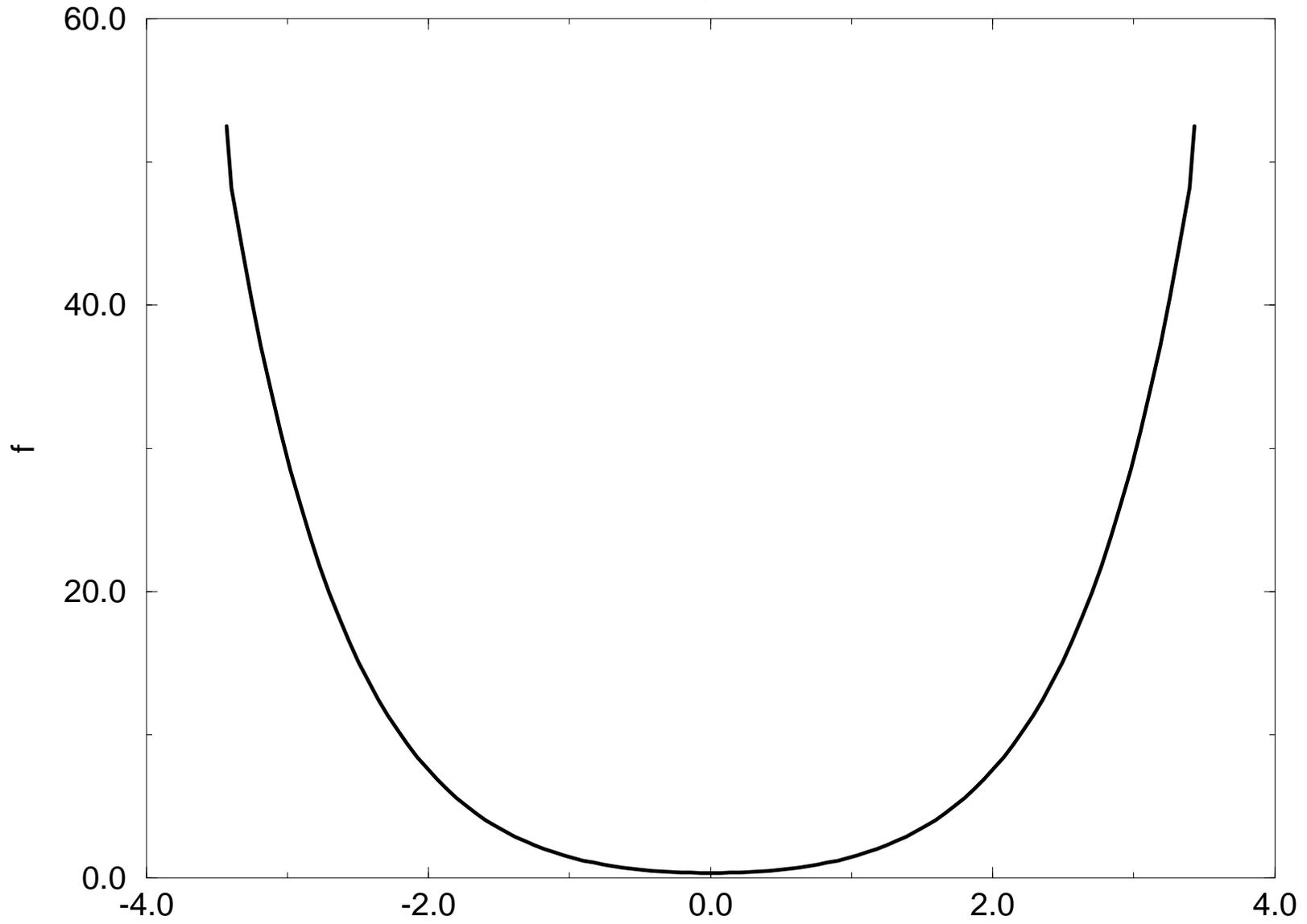
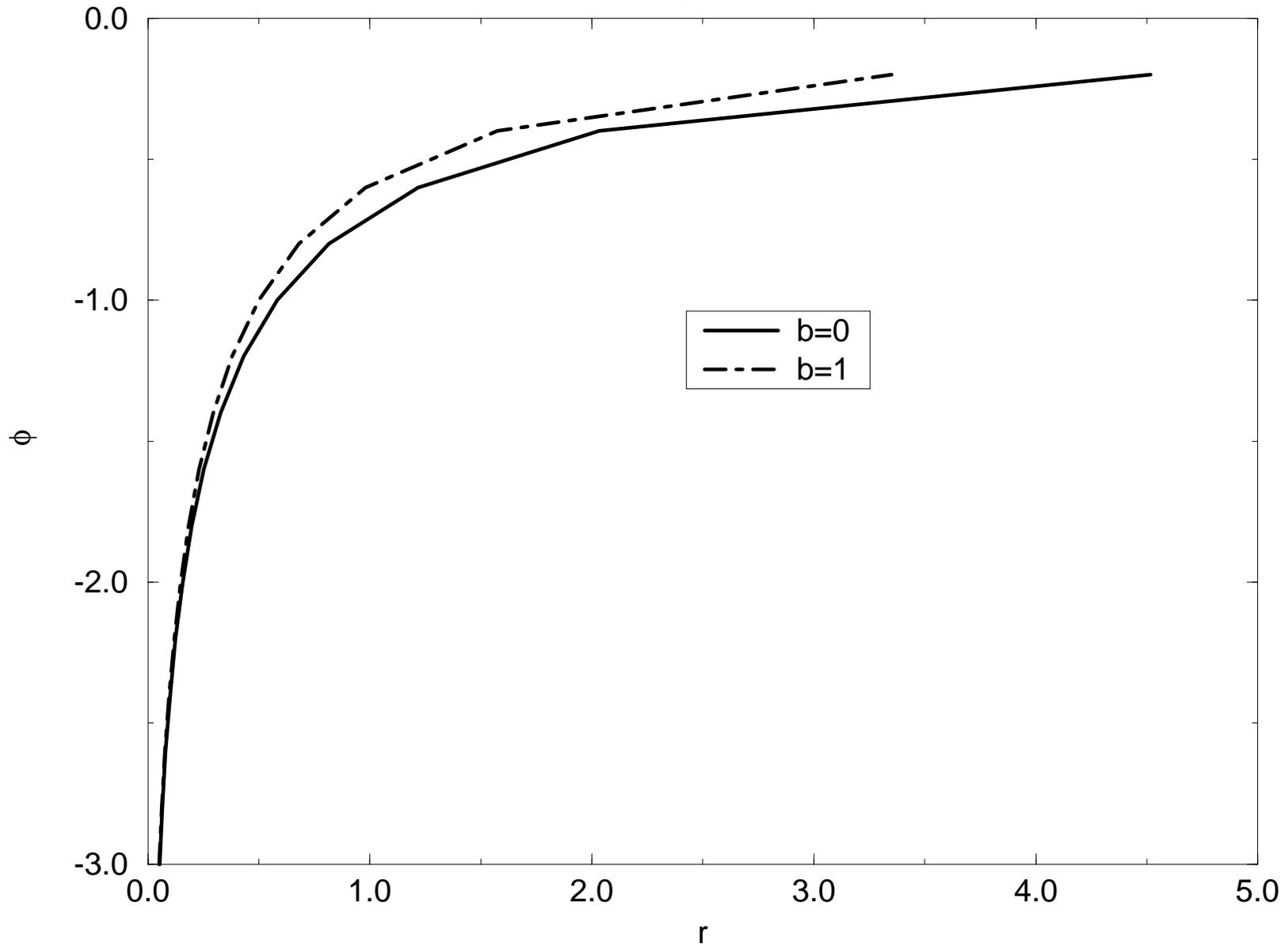
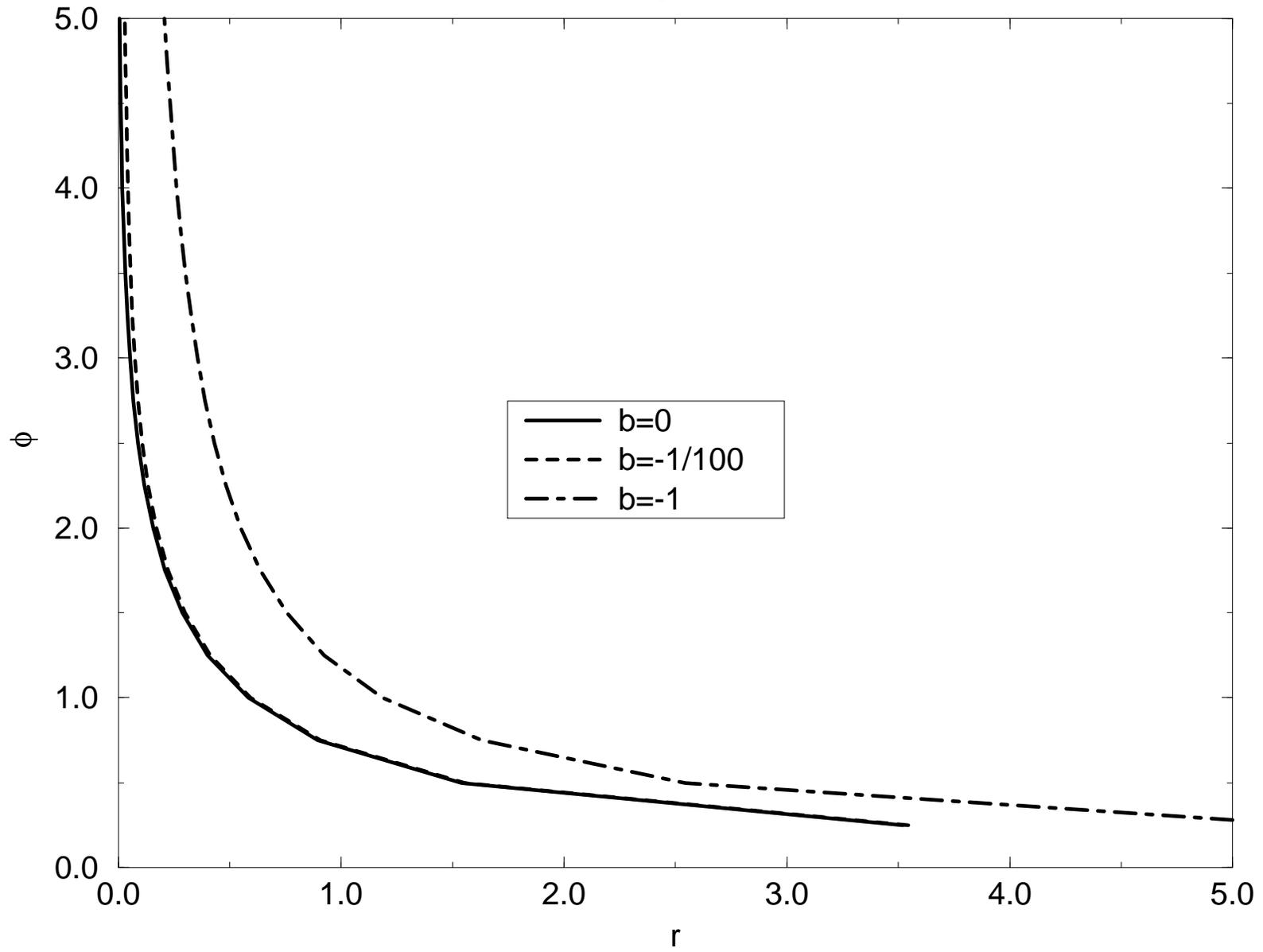
\end